\documentclass[preprint,review,12pt,authoryear]{elsarticle}

\usepackage{amssymb}
\usepackage{subfigure}
\usepackage{graphics}
\usepackage{amsmath}
\usepackage{lineno}

\journal{Powder Technology}

\begin{document}

\begin{frontmatter}



\title{Experiments and Discrete Element Simulation of the Dosing of Cohesive Powders in a Simplified Geometry}

\author[label1]{Olukayode I. Imole\corref{cor1}}
\ead{o.i.imole@utwente.nl}

\author[label1]{Dinant Krijgsman}
\author[label1]{Thomas Weinhart}
\author[label1]{Vanessa Magnanimo}
\author[label2]{Bruno E. Ch{\' a}vez Montes}
\author[label3,label4]{Marco Ramaioli}
\author[label1]{Stefan Luding}

\address[label1]{Multi Scale Mechanics (MSM), CTW, MESA+, University of Twente, P.O. Box 217, 7500 AE Enschede, The Netherlands.}
\address[label2]{Nestl{\' e} Product Technology Centre Orbe, Rte de Chavornay 3, CH-1350 Orbe, Switzerland.}
\address[label3]{Nestl{\' e} Research Center, Lausanne, Switzerland.}
\address[label4]{Department of Chemical and Process Engineering, FEPS (J2), University of Surrey, Guildford GU2 7XH, United Kingdom.}
\cortext[cor1]{Corresponding author}  %\\Email address: o.i.imole@utwente.nl
%\cortext[cor2]{o.i.imole@utwente.nl}

% \author{}
% 
% \address{}

\begin{abstract}
We perform experiments and discrete element simulations on the dosing of cohesive granular materials in a simplified geometry.
The setup is a simplified canister box where the powder is dosed out of the box through the action of a constant-pitch screw feeder connected to a motor. A dose consists of a rotation 
step followed by a period of rest before the next dosage. % From each experiment, the dosed mass per screw turn is recorded. 
% As found in several applications, the dosage is usually non-uniform due to the cohesive nature of the material sample. %A similar set-up was used in a previous study involving cohesionless grains [1].
% The challenge here is to use DEM simulations as a tool to gain insight into the micromechanical reasons for the behavior of the cohesive particles during the dosing action in order to ultimately optimize the dosing procedure.
%
From the experiments, we report on the operational performance of the dosing process through a variation of dosage time, coil pitch and initial powder mass. 
We find that the dosed mass shows an increasing linear dependence on the dosage time and rotation speed. In contrast, the mass output from the 
canister is not directly proportional to an increase/decrease in the number coils. By calibrating the interparticle friction and 
cohesion, we show that DEM simulation can quantitatively reproduce the experimental findings for smaller masses but also overestimate arching and blockage. With 
appropriate homogenization tools, further insights into microstructure and macroscopic fields can be obtained.

This work shows that particle scaling and the adaptation of particle properties is a viable approach to overcome the 
untreatable number of particles inherent in experiments with fine, cohesive powders and opens the gateway to simulating their flow in more complex geometries. % The confidence gained in this study, focusing on a simplified canister geometry, 
%paves the way to simulating the flow of these materials in more complex, real canister geometries, and to ultimately using numerical simulations as a virtual prototyping tool.

%and study the effects of particle size, size distribution and dosage time on the evolution of the dosed mass as function of time.

% The final step is to perform the micro-macro transition and provide a continuum description of the stress, density and velocity fields and thereafter provide a match/link to macroscopic design or optimization tools.

\end{abstract}


\end{frontmatter}


\newcommand{\dsmax} { \delta^{*}_\mathrm{max} }
\newcommand{\dmax} { \delta_\mathrm{max} }
\newcommand{\dmin} { \delta_\mathrm{min} }
\newcommand{\dnot} { \delta_\mathrm{0} }
\newcommand{\fhys} { f^{\mathrm{hys}} }
\newcommand{\fmin} { f_{\mathrm{min}} }
\newcommand{\rmic} { \rho^{\mathrm{mic}} }
\newcommand{\palp} { p_{\alpha }}
\newcommand{\valp} { V_{\alpha }}
\newcommand{\bt} { \beta_{t}}
\newcommand{\kc} { k_{c}}
\newcommand{\bkc} { K_{c}}

\newcommand{\kr} { k_{r}}
\newcommand{\bkr} { K_{r}}
\newcommand{\broll} { \beta_{\mathrm {roll}}}

\newcommand{\mwall} { \mu_{w}}
\newcommand{\bwall} { \beta_{\mathrm {wall}}}

\newcommand{\hktwo} {{\hat{ k}}_{2}}
\newcommand{\pf} { \phi_{f}}
\newcommand{\mtot} {m_{\mathrm{tot}}}
\newcommand{\mbin} {m_{\mathrm{bin}}}
\newcommand{\hini} {h_{\mathrm{ini}}}

\newcommand{\mdose} {m_{\mathrm{exp}}^{\mathrm{dose}}}
\newcommand{\ndose} {N_{\mathrm{exp}}^{\mathrm{dose}}}
\newcommand{\rhob} { \rho_{\mathrm{b}} }

%\begin{linenumbers}
\section{Introduction and Background}

The dynamic behavior of granular materials is of considerable interest in a wide range of industries (e.g. pharmaceutical, chemical and food processing). In these industries, every step in the product manufacturing process
contributes to the final quality of the product. Hence, if uniform product quality is to be achieved, a full understanding and control of the different stages of the production process is essential. In many applications, 
the transport and conveying of granular materials is a common process that forms a critical part of many production and delivery techniques. For example, transport to silos, process transport, controlled drug delivery and dosing 
of beverages all rely on an effective and uniform delivery of granular materials. 
Dosing consistently the correct amount of a soluble beverage powder is for instance the first step toward preparing a high quality beverage, but this process is also naturally 
conditioned by how the powder interacts with the water surface \cite{dupas2014powder}.
Also, the design of products for these processes is hugely dependent on having a good understanding of the transport behavior and metering process of granular assemblies.
%flow problems of powders arching, inconsistent dosed mass etc here...
 
When granular materials are being transported, the behavior of the granular material and the efficiency of the process will depend on several material properties including particle shape, particle size, surface roughness, frictional properties,
cohesion and moisture content among others. Discontinuities and inhomogeneities in the micro-mechanical behavior of bulk assemblies of granular materials are ever-present hence, changes in operating condition affect the flow behavior of granular
assemblies \cite{oda2000study}. Also the geometry of the transport media (boundary conditions) including wall friction and the loading/preparation procedure will play an important role. %Since granular materials show incrementally non-linear
%behavior, the direction of loading also plays an important role. Inhomogeneous distributions of stress and strain (anisotropy) is always present and leads to problems during transport \cite{wensrich2012measure} and difficulties 
%in predicting their flow behavior. 

Over the past decade, the mechanism during transport of granular materials have attracted significant interests and efforts from researchers. These efforts can be grouped into three classes namely, experimental, numerical modelling and developing
constitutive models to predict granular flows in conveying mechanisms \cite{yu1997theoretical, roberts1999influence}. The numerical modelling of granular flows has been based on %event driven (ED) simulations \cite{sebastian} and the more popular 
Discrete Element Method (DEM) as proposed in Ref.\ \cite{cundall1979discrete}.  
The earlier (more favored) experimental approach mostly involves the design and construction of experimental models of such applications followed by
series of studies and benchmark tests to determine quantities of interest and fine-tune the process to desirable levels. Thereafter, a scale-up of the process can be performed. 
In this case, the challenging task is the selection of 
relevant parameters and boundary conditions to fully characterize the flow rheology in these systems.

A measure of knowledge in characterizing dry, non-sticky powders exists. For example, rotating drum experiments and simulations to determine the dynamic angle of repose have 
been studied extensively as a means  to characterize non-cohesive powders \cite{Taberlet2006shape, brewster2009effects}. What has been less studied is the case where the powders 
are sticky, cohesive and less flowable like those relevant in the food industry. For these powders, 
dynamic tests are difficult to perform due to contact adhesion and clump formation. Inhomogeneities are also more rampant and flow prediction becomes even more troublesome.

Screw conveyors are generally used in process industries to transport bulk materials in a precise and steady manner. Materials like cereals, tablets, chemicals, pellets, 
salt and sand among others can be transported using screw conveyors. As simple as
this process may seem, problems of inaccurate metering, unsteady flow rates, bridging, channeling, arching, product inhomogeneity, segregation, high start up torques, equipment wear and variable 
residence time have been reported \cite{owen2009prediction,cleary2007modelling,owen2010screw, bortolamasi2001design}. In addition, the design and 
optimization of screw conveyors performance is not well understood and 
has been based on semi-empirical approach or experimental techniques using dynamic similarities as pointed out in Ref.\ \cite{bortolamasi2001design}. 
Earlier researchers have investigated the effect of various screw (auger) parameters including 
choke length (the distance beyond which the screw projects beyond the casing at the lower end of the intake) and pitch--diameter 
ratio (See Refs \cite {ghosh1967conveyance,stevens1962performance,roberts1962performance} and references therein).
Robert and Willis \cite{roberts1962performance} reported that since grain motion is largely influenced by its centrifugal inertia, augers with 
large diameters attain maximum output at lower speeds compared to those with small diameters. They also reported
that for maximum throughput during conveying, longer chokes are necessary.

The subject of modelling screw conveying of granular materials with the Discrete Element Method (DEM) \cite{cundall1979discrete} is fairly recent. 
One of the earliest work on this subject 
was reported in Ref.\ \cite{shimizu2001threedimensional} where the performance of horizontal and vertical screw conveyors are 
investigated and results are compared with empirical equations. 
In a related work, Owen et al. \cite{owen2009prediction} studied the performance of a long screw conveyor by introducing the so-called `periodic slice' model. 
Along this line, Cleary \cite{cleary2007modelling} investigated the 
effects of particle shape on flow out of hoppers and on the transport characteristics of screw conveyors. Experiments on the dosing of glass beads and cohesive
powders along with the discrete element simulation of the dosing of glass beads have also been reported \cite{ramaioli2007granular}.
A fundamental question is the extent to which discrete element simulations can predict the dosing of these powders, 
especially when the powders are cohesive.

In the current study, we use experiments and discrete element simulations to investigate the dosing of cohesive powders in a simplified canister geometry. 
The characterization of the experimental
samples, experimental set-up and procedure are presented in section \ref{sec:doseexp}. In section \ref{sec:numericalproc}, we present the force model, simulation parameters and homogenization
technique followed by a discussion of experimental and numerical results in section \ref{sec:discussion}. Finally, the summary, conclusions and outlook
are presented in section \ref{sec:doseconclusion}.

\section{Dosage Experiments}
\label{sec:doseexp}
In this section, we discuss in detail the experimental set-up and measurement procedure along with the material parameters of the experimental sample.

\subsection{Sample Description and Characterization}
\label{sec:material}

The cohesive granular sample used in this work is cocoa powder with material properties shown in Table \ref{tablematerialpptaa}. %and the morphology shown in Fig.\ \ref{materialsem}
The particle size distribution (PSD) is obtained by the ``dry dispersion module'' of the Malvern Mastersizer 2000 (Malvern Instruments Ltd., UK), while the particle density is obtained by helium pycnometry (Accupyc, Micromeritics, US).
The span is defined as the width of the distribution based on the 10\%, 50\% and 90\% quantile.
The experiments were performed over a relatively short period under ambient conditions and samples are sealed in air-tight bags when not in use to minimize effects that could arise due to
changes in the product humidity.

\begin{table}[!htbp]
\centering
 % \begin{tabular}{ cp{3.0cm}p{7.0cm}p{1.9cm} }
 \begin{tabular}{lrll}
    \hline
     \bf{Property} &  & \bf{Unit} & \bf{Value} \\ [0.4ex]
    \hline %\\ [0.4ex]
    Size distribution & $d_{10}$& $\mu$m & 31.55\\ [0.4ex]
    %\hline
     &  $d_{50}$ & $\mu$m & 184\\ [0.4ex]
    %\hline
    & $d_{90}$ & $\mu$m  & 979.19 \\ [0.4ex]
    %\hline
    Span & $(d_{90} - d_{10})/d_{50}$  & [-] & 5.151 \\ [0.4ex]
    %\hline
     Particle Density &  &[kg/m\textsuperscript{3}] & 1427\\ [0.4ex]
    %\hline
    Specific surface area &  & m$^2$/g & 0.088\\ [0.4ex]
    %\hline
%     Moisture content & & \% & 5.68 \\ [0.4ex]
    \hline
  \end{tabular}
  \caption{Material properties of the experimental cocoa sample.
%SL why? ,imole2013hydrostatic}.}
}
  \label{tablematerialpptaa}
\end{table}

\subsection{Experimental Set-up}
\label{sec:expsetuprpoc}

\begin{figure} [!ht]
\centering
\subfigure[]{\includegraphics[scale=0.30]{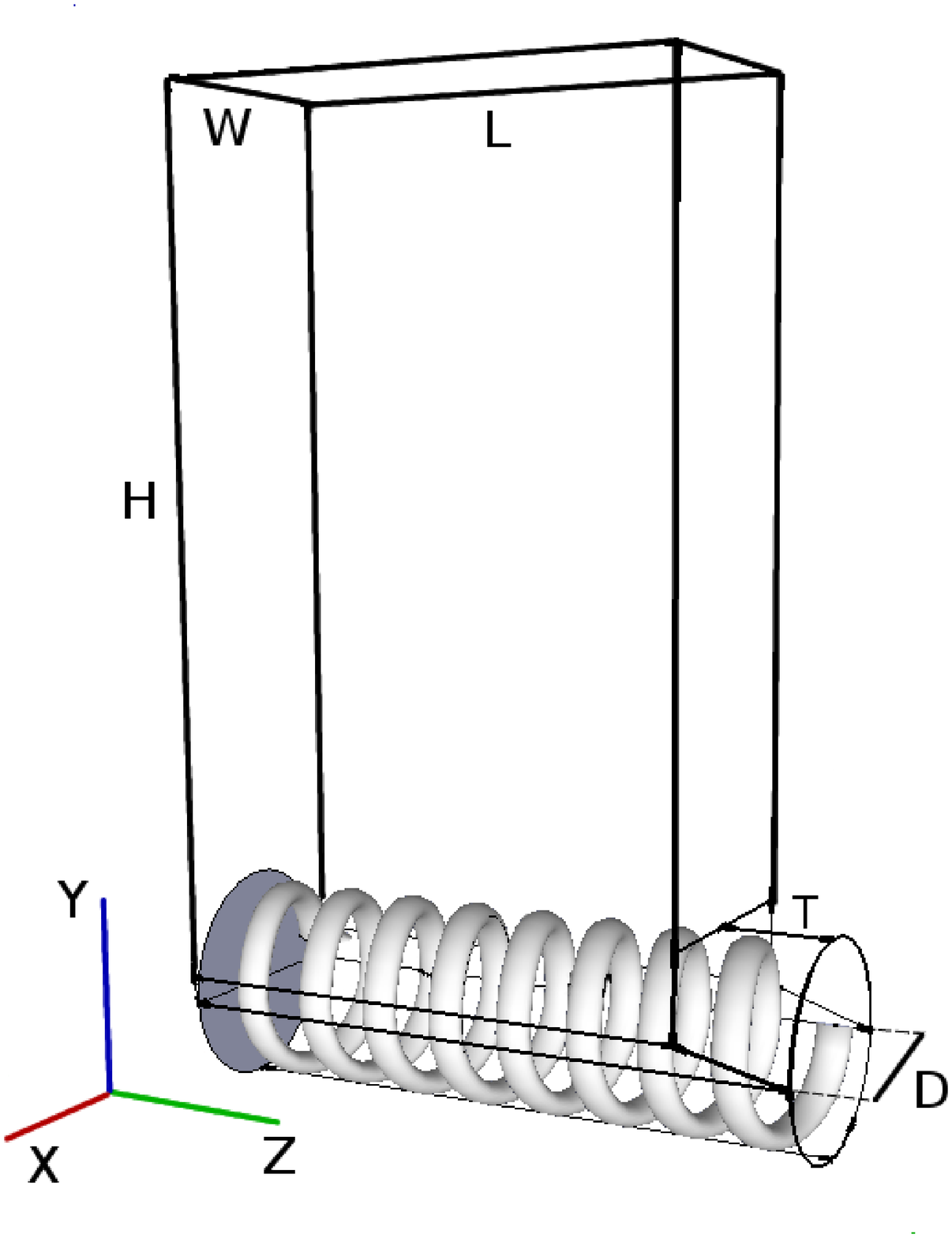}\label{canister}}
\subfigure[]{\includegraphics[scale=0.30]{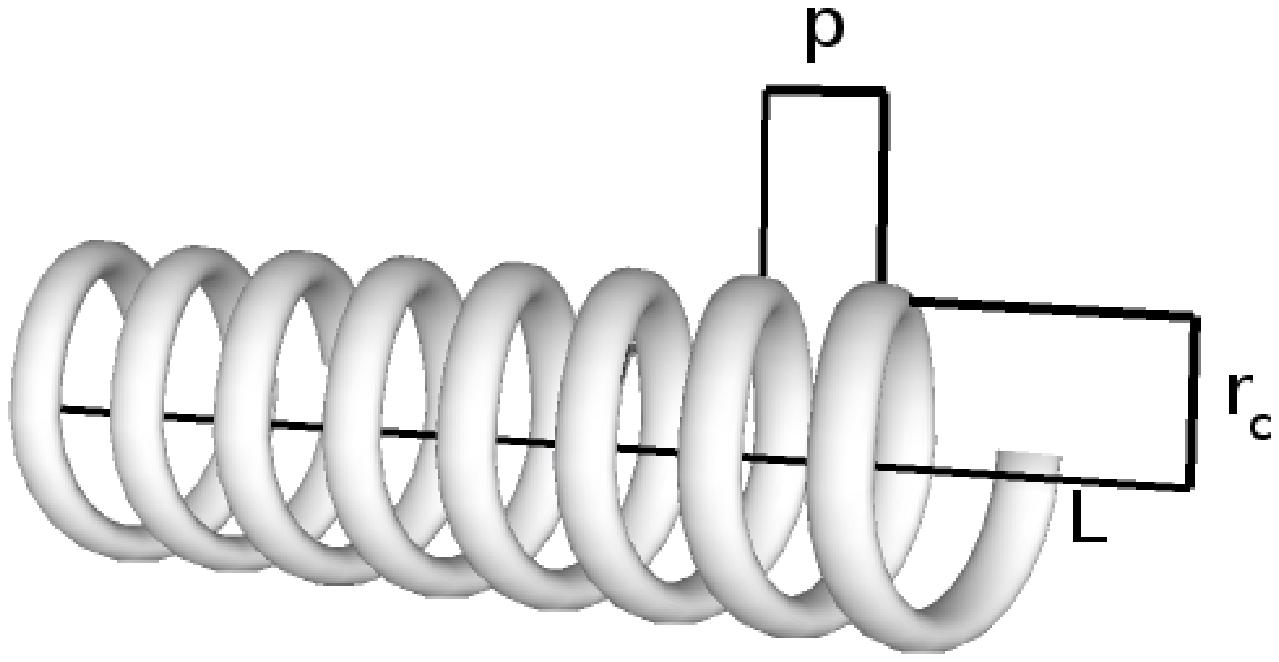}\label{coil}}
\caption{Schematic representation of the (a) simplified canister, (b) the coil used for the dosing experiments and simulation, with box length ($L$), width ($W$), height($H$), throat 
length $T$, outlet diameter ($D$), coil radius $r_{c}$ and pitch $p$.}
\label{setupschematic}
\end{figure}

The setup is a simplified canister box where the powder is dosed out of the box through the action of a constant-pitch screw feeder connected to a motor. A schematic representation of the experimental set-up is shown in 
Fig.\ \ref{setupschematic} along with the dimensions on Table \ref{parametertablea}. A typical experiment begins with the careful filling of the canister with the exit closed until a pre-determined powder mass is reached.
Care is taken to ensure that the initial profile of the powder surface is as flat as possible and that any pre-compaction that may arise due to shaking or vibrations are minimized.
Subsequently, the dosing experiment begins with the rotation of the screw for a specified time duration followed by an intermediate rest before the next dosage. 
The dosed mass per screw turn is recorded through a weighing scale connected to a computer. The experiment is complete when the cumulative dosed mass recorded for three consecutive doses is less than 0.15grams indicating
either the box is empty or the powder is blocked through arching in the canister. In addition, to obtain and post-process the profiles of the sample surface during the experiments, 
an external camera (Logitech HD Pro, Logitech Int'l SA) was mounted in front of the canister box and a 
video recording of each experiment was obtained. 

%%%%%%%%%%%%%%%%%%%%%%%%%%%%%%%%%%%%%%%%%%%%%%%%%%%%
\begin{table}
\centering
 % \begin{tabular}{ cp{3.0cm}p{7.0cm}p{1.9cm} }
 \begin{tabular}{lr}
    \hline
    \bf{Parameter} & \bf{Value}  \\ [0.4ex] %& \bf{Simulation}  \\ [0.4ex]
    \hline %\\%[-0.4ex]
    Canister dimensions \\($L \times W \times H$) & $60 \times 23 \times 170$ mm \\ [0.4ex] %& $60 \times 22 \times 120$ [mm] \\ [0.4ex]
    %\hline
    Throat length ($T$) & 10 mm \\ [0.4ex] %& 10 [mm] \\ [0.4ex]
    %\hline
    Outlet diameter ($D$) & 23 mm \\ [0.4ex] %& 22 [mm] \\ [0.4ex]
    %\hline
    Coil thickness & 2 mm \\ [0.4ex] %& 2 [mm]  
    %\hline
    Coil length & 70 mm \\ [0.4ex] %& 70 [mm]  
    %\hline
    Coil radius ($r_c$) & 10.4 mm \\ [0.4ex] %& 70 [mm]  
    %\hline
    Number of coils & 4 (Wide), 8 (Narrow) \\ [0.4ex] %& 4 (Wide), 8 (Narrow)  
    %\hline
    Coil pitch & 17.5 mm (Wide), 8.75 mm (Narrow) \\ [0.4ex] %& 4 (Wide), 8 (Narrow)  
    %\hline
    Coil angular velocity ($\Omega$) & 90 rpm (9.42 rad/s)\\ [0.4ex] %& 4 (Wide), 8 (Narrow)  
  \hline
  \end{tabular}
  \caption{Summary of system parameters used 
in the experiments and DEM simulations.
}
  \label{parametertablea}
\end{table}

\subsubsection{Image Processing}
\label{sec:imageprocess}

Snapshots of the profile of the powder surface during each experiments were obtained using a camera attached to the experimental set-up. To improve the quality of the snapshots and for comparison, we use the open-source
software FIJI \cite{schindelin2012fiji} to post-process the images following a three step procedure, namely quality adjustment, binarization and extraction of the lateral surface of the powder. 
In the first stage, we adjust the quality of the images 
by first selecting the region of interest and enhancing its contrast. In the second stage, the image is binarized into black (0) and white (1) pixels such that the area containing the bulk sample is easily differentiated from
other areas in the picture. In the final step, we iteratively move along the length of the image from top to bottom to trace out the profile of the powder surface.

For a given rotation speed, the linear coil (push) velocity is:
\begin{equation}
 V_{z}= \frac{p\omega}{2\pi}
\label{eq:linvel}
\end{equation}
where $\omega$ is the angular velocity of the coil and $p$ is the pitch of the coil. Also, the coil tangential velocity is:
\begin{equation}
  V_{t}= \omega r_{c}
\label{eq:tangvel} 
\end{equation}
where $r_{c}$ is the coil radius. Accordingly, the expected dosed mass for a single rotation of the coil is:
\begin{equation}
 \mdose = \rhob \cdot V_c \cdot n_{t} = \rhob \cdot p \cdot \pi {r_c}^2 \cdot n_{t}
\label{eq:mdoseexp}
\end{equation}
where $\rhob$ is the bulk density, $V_c$  is the volume within a single pitch and $n_t =  t_{d} \cdot \omega/2\pi$ is the number of rotations completed within a given dosing time $t_{d}$. The expected number of doses is then:

\begin{equation}
 \ndose = \frac{\mtot}{\mdose}
\label{eq:ndoseexp}
\end{equation}
where $\mtot$ is the total initial mass filled into the canister.

\section{Numerical Simulation}
\label{sec:numericalproc}

The numerical simulation was carried out using the open source discrete element code MercuryDPM \cite{thornton2012modeling, weinhart2012from}. 
Since DEM is otherwise a standard method, only the contact model and the basic system parameters are briefly discussed.

\subsection{Force Model}
\label{sec:forcemodel}

\begin{figure*}[!ht]
 \centering
\subfigure[]{\includegraphics[scale=0.30]{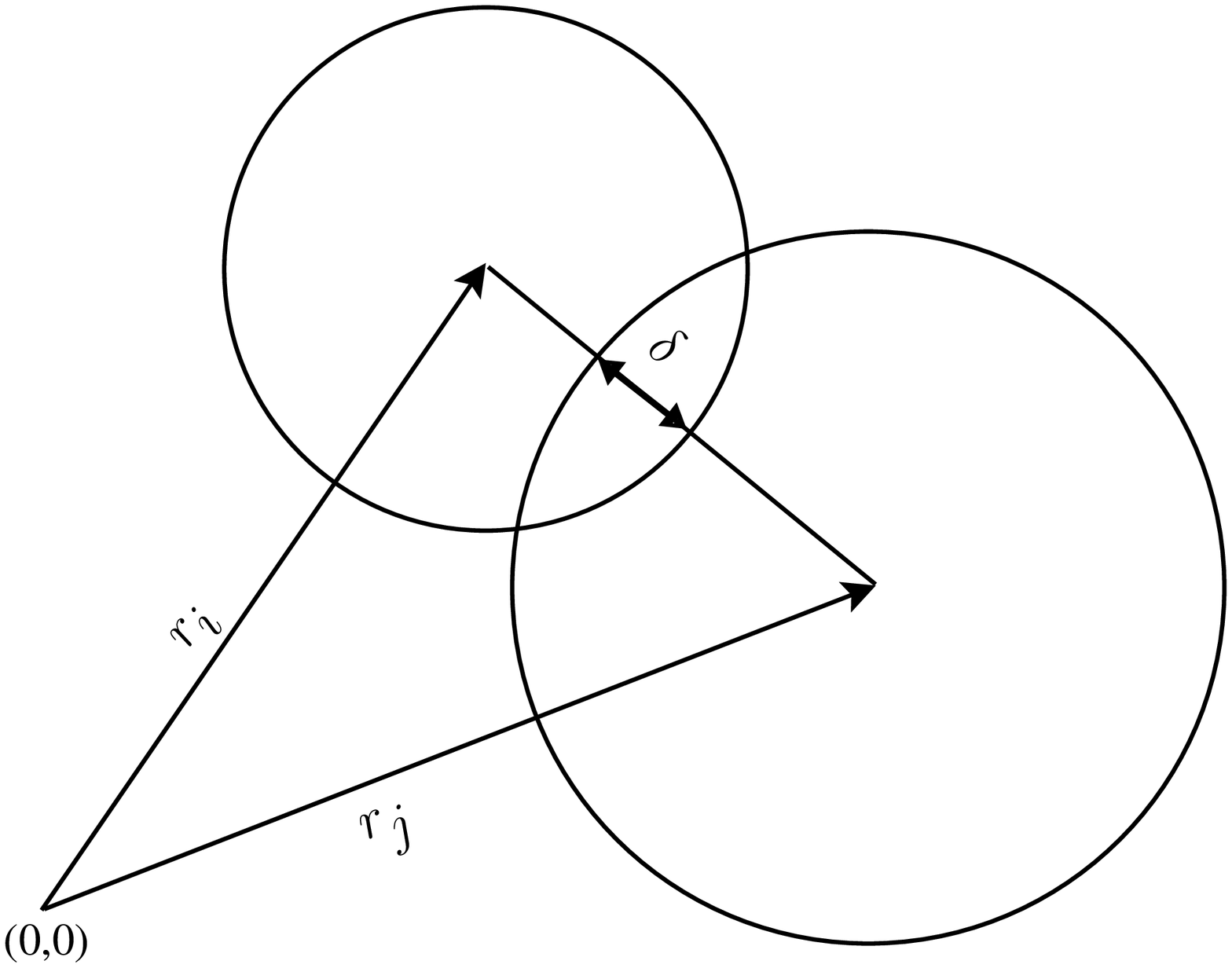}\label{overlap}}
\subfigure[]{\includegraphics[scale=0.40]{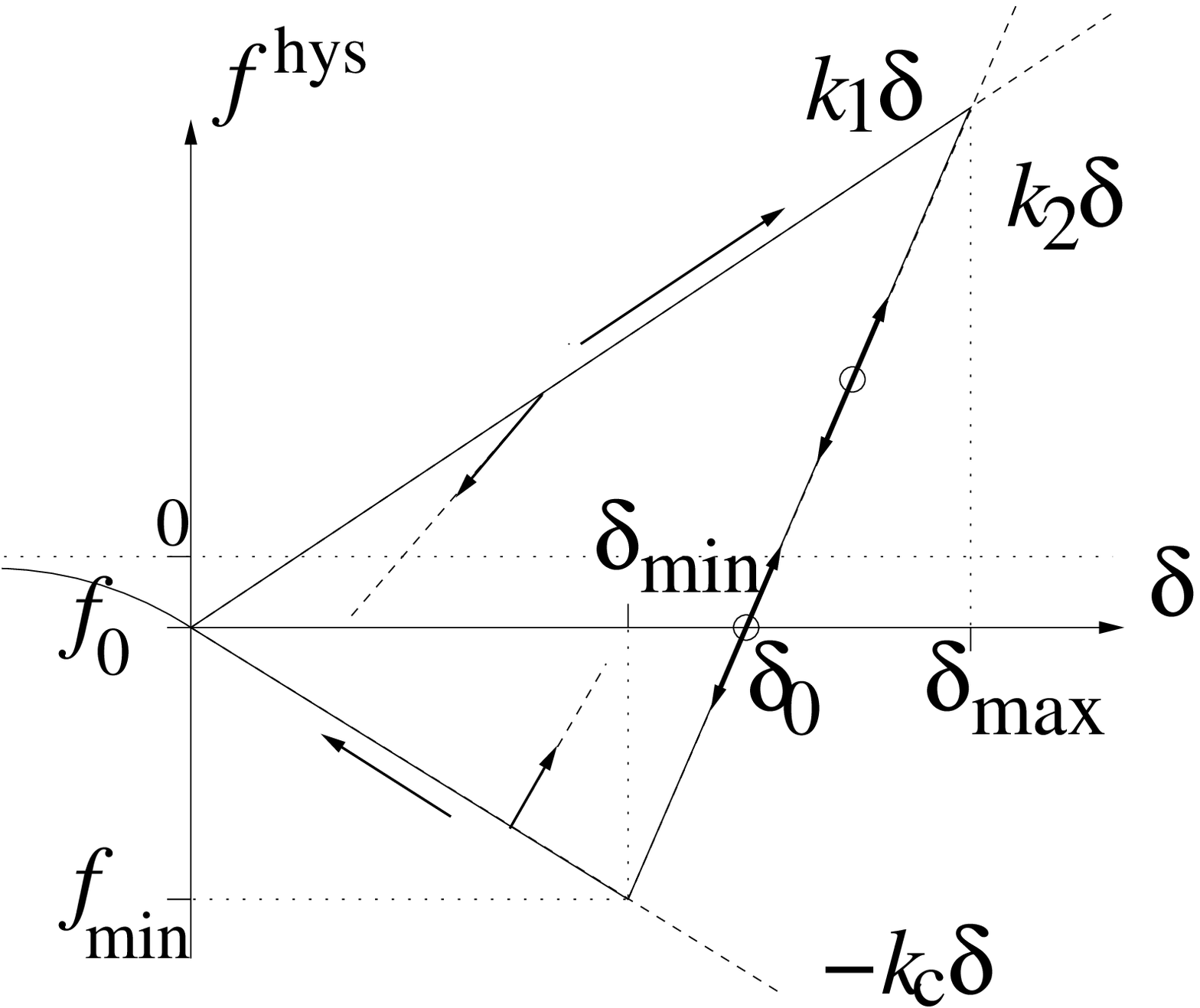}\label{model}}\\
\caption{(a) Two particle contact with overlap $\delta$ in normal direction. (b) Schematic graph of the linear,
hysteretic, adhesive force-displacement model in normal direction} %Arrows pointing right
\label{contactmodel}
\end{figure*}

Since realistic and detailed modeling of the deformations of particles in contact with each other is much too complicated, we relate the interaction force to the
overlap $\delta$ of two particles as shown in Fig.\ \ref{overlap}. Thus, the results presented here are of the same quality as the simplifying assumptions about the force-overlap relations made.
However, it is the only way to model larger samples of particles with a minimal complexity of the contact properties, taking into account the relevant phenomena: non-linear contact elasticity, plastic
deformation, and adhesion.

In this work, we use the Luding's linear hysteretic spring model \cite{luding2008cohesive} -- which is a simplified version of more complicated non-linear hysteretic force laws 
\cite{walton1986viscosity, zhu1991prediction, sadd1993contact, tomas2000particle, tomas2004fundamentals}. The adhesive, plastic (hysteretic) normal force is given as:

\begin{equation}
f^{\mathrm{hys}} = \left\{ 
  \begin{array}{l l}
    k_1 \delta & \quad \text{if $k_2 (\delta - \dnot) \ge  k_1 \delta$ }\\
    k_2 (\delta - \dnot) & \quad \text{if $k_1 \delta > k_2 (\delta - \dnot) > -k_c \delta$}\\
    -k_c \delta  & \quad \text{if $-k_c \delta \ge k_2 (\delta - \dnot) $}
  \end{array} \right.%\]
\label{eq:forcemodel}
\end{equation}
with $k_1 \le k_2 \le \hat{k}_2$ as shown in Fig.\ \ref{model} where $\hat{k}_2$ is the maximum stiffness and $f_0$ has been set to zero. 
During initial loading the force increases linearly with the overlap, until the maximum overlap $\dmax$ is reached ($\dmax$ is kept in memory as a history variable). The line
with slope $k_1$ thus defines the maximum force possible for a given $\delta$.

During unloading the force drops on a line with slope $k_2$ , which depends, in general, on $\dmax$. The force at $\delta = \dmax$ decreases to zero, at overlap $\dnot = (1 - k_1 /k_2 )\dmax$ , which resembles the plastic contact 
deformation. Reloading at any instant leads to an increase of the force along the same line with slope $k_2$, until the maximum force is reached; for still increasing $\delta$, the force
follows again the line with slope $k_1$ and $\dmax$ has to be adjusted accordingly \cite{luding2008cohesive}.

Unloading below $\dnot$ leads to attractive adhesion forces until the minimum force $-k_c \dmin$ is reached at the overlap $\dmin = (k_2 - k_1 )\dmax /(k_2 + k_c )$, a function of the
model parameters $k_1, k_2, k_c$, and the history parameter $\dmax$. Further unloading leads to attractive forces $\fhys = - k_c \delta$ on the adhesive branch. The highest
possible attractive force, for given $k_1$ and $k_2$ , is reached for $k_c \to \infty $, so that one has $\fmin \ge -(k_2  - k_1 )\dmax$ for arbitrary $k_c$.

A more realistic behavior will be a non-linear un-/re-loading behavior. However, due to a lack of detailed experimental
information, the piece-wise linear model is used as a compromise. One reasonable refinement, which accounts for an {\em{increasing unloading stiffness with deformation}}, is a $k_2$
value dependent on the maximum overlap. This also implies relatively small and large plastic deformations for weak and strong contact forces, respectively. Unless a constant
$k_2 = \hktwo$ is used, the contact model \cite{luding2005discrete, luding2007contact, luding2006about}, requires an additional quantity, i.e., the plastic flow limit overlap

\begin{equation}
 \dsmax = \frac{\hktwo}{\hktwo - k_1} \pf \frac{2a_1a_2}{a_1 + a_2},
\label{eq:pflimit}
\end{equation}
with the dimensionless plasticity depth $\pf$, defined relative to the reduced radius. If the overlap is larger than a $\pf$
fraction of the particle radius (for radius $a_1 = a_2$), the (maximal) constant stiffness constant stiffness $\hktwo$ is used.
For different particle radii, the reduced radius increases towards the diameter of the smaller particles in the extreme case of 
particle-wall contacts (where the wall-radius is assumed infinite).

Note that a limit stiffness $k_2 \le \hktwo$ is desirable for practical reasons.
If $k_2$ would not be limited, the contact duration could become very small so that the time step
would have to be reduced below reasonable values. For overlaps smaller than
$\dsmax$, the function $k_2(\dmax)$ interpolates linearly between $k_1$ and $k_2$:

\begin{equation}
k_2(\dmax) = \left\{ 
  \begin{array}{l l}
     \hktwo & \quad \text{if $\dmax \ge \dsmax $ }\\
    k_1 + (\hktwo - k_1)\frac{\dmax}{\dsmax}& \quad \text{if $\dmax < \dsmax$}. \\
%     -k_c \delta  & \quad \text{if $-k_c \delta \ge k_2 (\delta - \dnot) $}
  \end{array} \right.%\]
\label{eq:ktwodmax}
\end{equation}

The implementation of the tangential forces and torques have been described extensively 
in Refs.\ \cite{luding2005discrete, luding2007contact, luding2006about, luding2008cohesive}.

\subsection{Simulation Procedure and Parameters}
\label{sec:simparameters}

The actual number of particles present in the bulk powder used in the dosing experiments are of the order several billions.
The realistic simulation of the exact size is exceptionally challenging due to computational cost. Due to this constraint, one choice available 
is to either scale the system size while keeping particle properties fixed. The other choice is to do the contrary,  namely keeping the system size fixed and scaling (or coarse graining)
the particle sizes up by essentially making them meso particles. We choose to do the latter, namely using the same system size in simulation as in experiments and  increasing the size of our 
particles so that each meso-particle in our system can be seen as an ensemble of smaller constituent particles. The system parameters used in both simulation and experiment are presented in 
Table \ref{parametertablea}. Note that the number of coils refer to the number of turns in the coil which, when divided by the length of the coil should give an indication of the pitch of the coil.
Typical numerical parameters used in the DEM simulation are listed in Table \ref{parametertableb}.

\begin{table}
\centering
 % \begin{tabular}{ cp{3.0cm}p{7.0cm}p{1.9cm} }
 \begin{tabular}{lrl}
    \hline
    \bf{Parameter} & \bf{Experiment} & \bf{Simulation}  \\ [0.4ex]
    \hline \\[-0.4ex]
    %\hline
    Number of Particles  &  $ > 10^{10}$ & 3360 for $\mtot = 48$ grams   \\ [0.4ex]
    %\hline
    Mean particle diameter ($\langle d \rangle$) & 0.184mm &  2.50 mm  \\ [0.4ex]
    %\hline
    Particle density ($\rho$) & $1.427 \times 10^{-6}$ kg/mm$^3$ & $1.427 \times 10^{-6}$ kg/mm$^3$ \\ [0.4ex]
    %\hline
    Polydispersity ($w$) & see Table \ref{tablematerialpptaa} & $r_{\mathrm {max}}/r_{\mathrm {min}} = 3$   \\ [0.4ex]
    %\hline
    Restitution coefficient ($e$) &  [--]&  0.45  \\ [0.4ex]
    %\hline
    Plasticity depth ($\pf$) &  [--]&  0.05  \\ [0.4ex]
    %\hline
    Maximal elastic stiffness ($k = \hktwo$) &  [--]&  24067 kg/s$^2$  \\ [0.4ex]
    %\hline
    Plastic stiffness ($k_1/k$) &  [--]&  5  \\ [0.4ex]
    %\hline
    Cohesive stiffness ($k_c/k$) &  [--]& 0.873 (varied 0--1) \\ [0.4ex]
    %\hline
    Friction stiffness ($k_t/k$) &  [--]&  0.286  \\ [0.4ex]
    %\hline
    Rolling stiffness ($k_r/k$) &  [--]&  0.286  \\ [0.4ex]
    %\hline
    Coulomb friction coefficient ($\mu$) &  [--]&  0.5  (varied 0.5--0.65)\\ [0.4ex]
    %\hline
    Rolling friction coefficient ($\mu_r$) &  [--]&  0.5  \\ [0.4ex]
    %\hline
    Normal viscosity ($\gamma_n = \gamma$) &  [--]&  0.0827 kg/s \\ [0.4ex]
    %\hline
    Friction viscosity ($\gamma_t/\gamma$) &  [--]&  0.286 \\ [0.4ex]
    %\hline
    Rolling viscosity ($\gamma_r/\gamma$) &  [--]&  0.286 \\ [0.4ex]
    %\hline
    Wall friction ($\mu_w$) &  [--]&  0.2 \\ [0.4ex]
    %\hline
    Contact duration $t_c$ &  [--]&  $ 1.1297 \times 10^{-4}$ s  \\ [0.4ex]
    %\hline
  \hline
  \end{tabular}
  \caption{Numerical values of parameters used in experiment and DEM simulations. }
  \label{parametertableb}
\end{table}

The numerical implementation of the dosing test is as follows. The particles are generated and positioned on regular grid points within the dimensions of the box. To avoid any initial overlap of particles,
either with the coil, surrounding wall or with  other particles, we ensure that the initial position of the lowest particle during this generation stage is higher than the diameter of the coil. %Also, we  position 
%sufficient number of particles along the width and length of the box such that only maximum diameter of the particles
Subsequently, the particles are allowed to fall under gravity and are left to settle and dissipate their energies for 2 seconds while the coil is not rotating. 
We find that for strong cohesion, 
this preparation method leads to initial inhomogeneities and irregular packing within the circumferential area of the coil during the settling phase. This gives rise to irregular dose
patterns and increases the possibility of arches (blockage) forming just above the screw. To minimize this, the particles are allowed to settle with a initial cohesive stiffness $k_c/k=0.3$ such that the
initial packing structure is homogeneous while the actual cohesion is activated after the settling phase.

\subsection{Homogenization Technique}
\label{sec:avgingtech}

In order to drive macroscopic fields such as density, velocity and stress tensor from averages of the microscopic discrete element variables such as the positions, velocities and forces of the constituent particles,
we use the coarse-graining method proposed in Refs.\ \cite{goldhirsch2010stress,weinhart2012from,weinhart2012closure,weinhart2013coarse}.

The microscopic mass density of a flow at a point $\mathbf{r}_{\alpha}$ at time $t$ is defined by 

\begin{equation}
 \rmic({\bf{r}},t) = \sum_{i=1}^{N} m_i \delta ({\bf{r}}-{\bf{r}}_i(t)),
\end{equation}
where $\delta({\bf{r}})$ is the Dirac delta function and $m_i$ and $r_i$ are the mass and center of mass position of particle $i$. Accordingly, the macroscopic density can be defined as:

\begin{equation}
 \rho({\bf{r}},t) = \sum_{i=1}^{N} m_i \mathcal{W} ({\bf{r}}-{\bf{r}}_i(t)),
\end{equation}
where the Dirac delta function has been replaced with an integrable `coarse-graining' function $\mathcal{W}$ whose integral over the domain is unity and has a predetermined width, or homogenization scale $w$.
In this work, we use a Gaussian coarse-graining function.

The homogenized momentum density is also defined as:

\begin{equation}
 \palp({\bf{r}},t) = \sum_{i=1}^{N} m_i v_{i\alpha} \mathcal{W}({\bf{r}}-{\bf{r}}_i).
\end{equation}
with $v_{i\alpha}$ the velocity of particle $i$. The macroscopic velocity field  $\valp({\bf{r}},t)$ is defined as the ratio of momentum and density fields, $\valp({\bf{r}},t) = \palp({\bf{r}},t)/\rho({\bf{r}},t)$. 
Comparing other fields, like stress- and structure- tensors as shown in Refs. \cite{goldhirsch2010stress,weinhart2012from,weinhart2012closure,weinhart2013coarse}, is beyond the scope of this study.

%%%%%%%%%%%%%%%%%%%%%%%%%%%%%%%%%%%%%%%%%%%%%%%%%%%%%new definition of height
In order to obtain the height variation during the dosing process $h_z$  as in experiment, we average over the height and the depth of the drum 

The height variation of the packing during the dosing is given as:

\begin{equation}
 h_{z} = \frac{\mbin (z,t)}{\mbin(z,0)} \cdot \hini = \frac{\rho (z,t)}{\rho(z,0)} \cdot \hini,
\end{equation}
assuming an almost constant bulk density.
$\mbin(z,t)$ is the mass change as function of time during the dosing process, $\mbin(z,0)$ is the initial mass of the particles at time $t =$ 0  and $\hini$ is the initial height of the packing. Furthermore, 
the mass in a bin as function of time is:

\begin{equation}
 \mbin (z,t) = \int_0^{\Delta z} \int_0^{H} \int_0^{D} \rho (x,y,z,t) dx dy dz
\end{equation}
where $D$ is the depth (or width) of the drum, $H$ the box height, and $\Delta z$ the bin width.
%%%%%%%%%%%%%%%%%%%%%%%%%%%%%%%%%%%%%%%%%%%%%%%%%%%%%%%%%%%%%%%%%
%
\section{Experiments}
\label{sec:discussion}

In this section, we present the results from the experiments and simulations and their comparison. %In section \ref{sec:expdiscussion}, we present experimental results on a 
%variation of different system parameters. 
For an understanding of the dosing process, in the following, we present experimental results on the effect of initial mass, number of coils and dosage time.

\subsection{Effect of Initial Mass in the Canister}
\label{sec:inimass}

\begin{figure*}[!ht]
 \centering
\subfigure[]{\includegraphics[scale=0.40, angle=270]{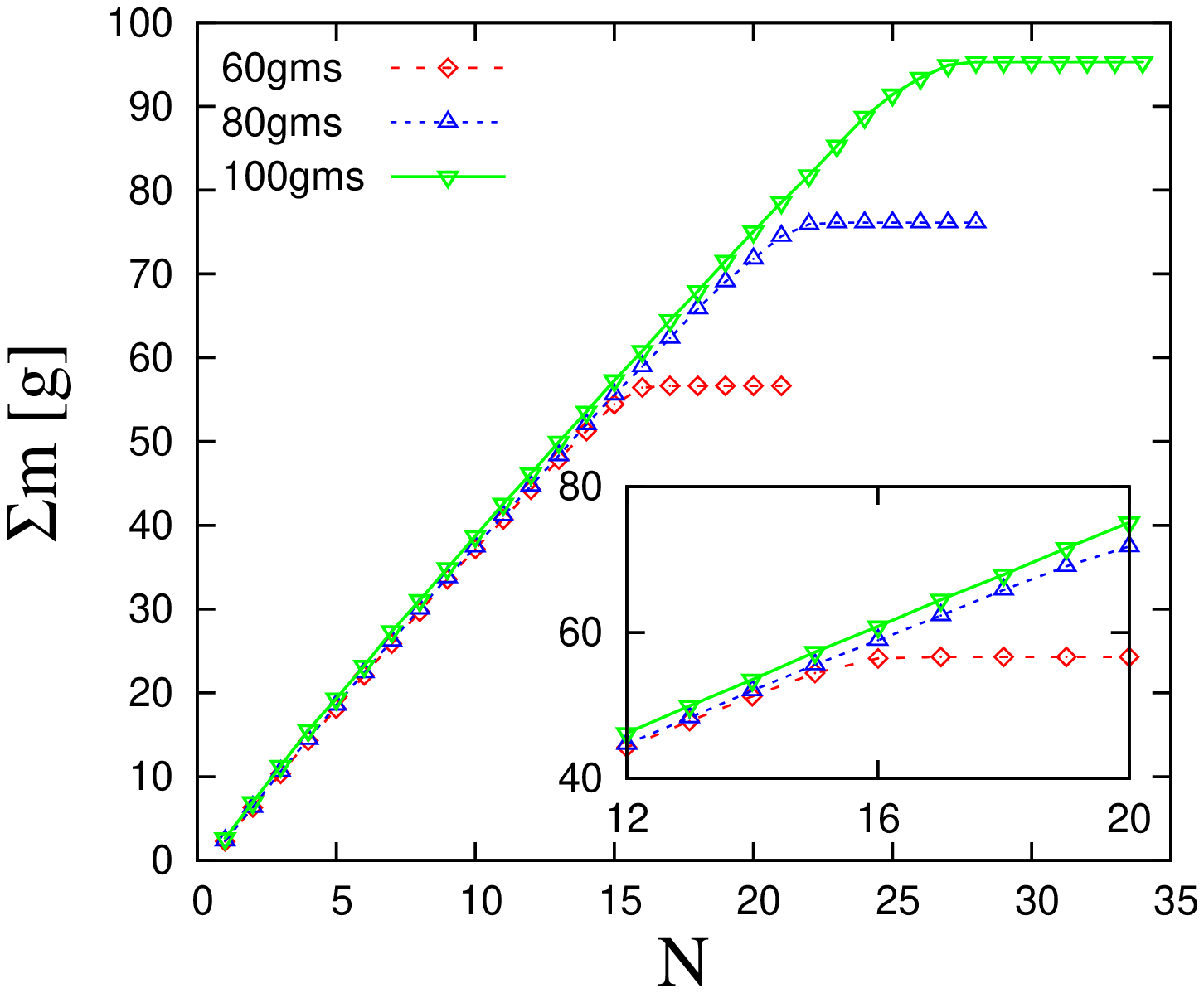}\label{massres}}
\subfigure[]{\includegraphics[scale=0.40,angle=270]{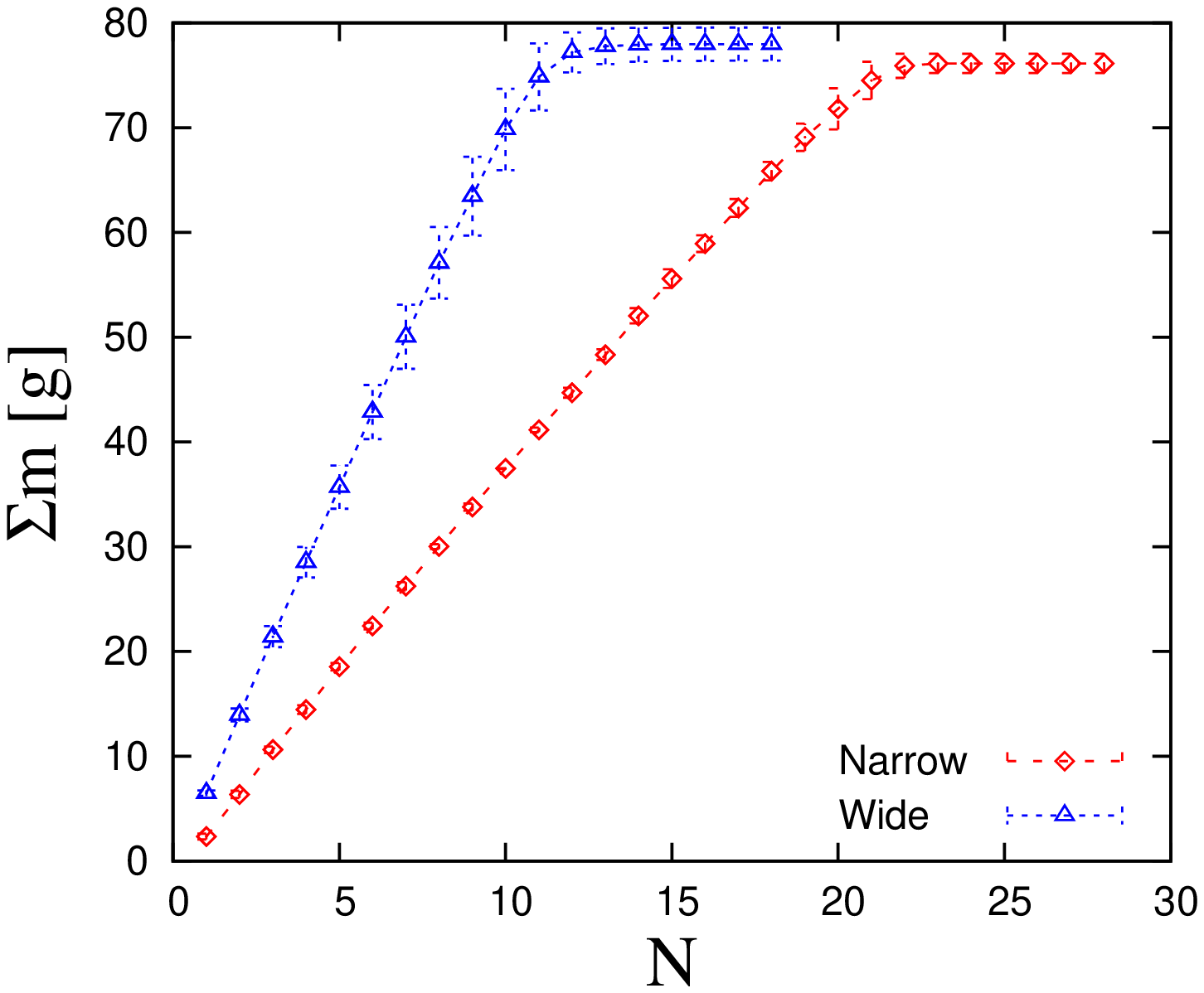}\label{screwres}}\\
\caption{(a) Cumulative dosed mass from experiments plotted as function of the number of doses for (a) different initial mass in the box (b) different coils (pitch).} %Arrows pointing right
\label{numbeofdosepararams}
\end{figure*}

In Fig.\ \ref{massres}, we plot the cumulative dosed mass as function of the number of doses for sample masses $\mtot=$ 60g, 80g and 100g in the canister. 
For these experiments, a dose consists of the rotation of the narrow pitch screw
for 2 seconds at a speed of 90rpm. As expected the number of doses increases with increasing the mass of powder filled in the canister. 
The number of doses recorded when 99 percent of the total powder mass in the canister is dosed are 16, 23 and 30, for the 60, 80 and 100g fill masses respectively.

The mass per dose obtained for different initial masses is close as shown by the near collapse of the data on each other.
We however note that the sensitivity of the mass per dose to the initial mass is tiny as seen in the inset.

\subsection{Effect of Number of Coils}
\label{sec:nocoil}

In Fig.\ \ref{screwres}, we plot the cumulative dosed mass as function of the number of doses for experiments with two different coils namely a wide coil with 4 coils (or crests) and a narrow
coil with 8 crests. The initial powder mass in the canister is 80g and a dose consists of the rotation of the coil for 2 seconds at a speed of 90rpm. The error bars represent the standard deviation over three
experimental runs for each test. From Fig.\ \ref{screwres}, it is evident that the dosed mass per coil turn for the coil with the wide pitch is higher compared to the dosed mass reported for the narrow screw. As a result, the
cumulative dosed mass recorded for the coil with the wide pitch increases faster (with slope 7.15 g/dose) in comparison to the narrow one (3.705 g/dose). This indicates that increasing the number of coils from 4 to 8 leads to
almost double increase in the mass per dose.

\subsection{Effect of Dosage Time}
\label{sec:dosagetime}

\begin{figure*}[!ht]
 \centering
\subfigure[]{\includegraphics[scale=0.28, angle=270]{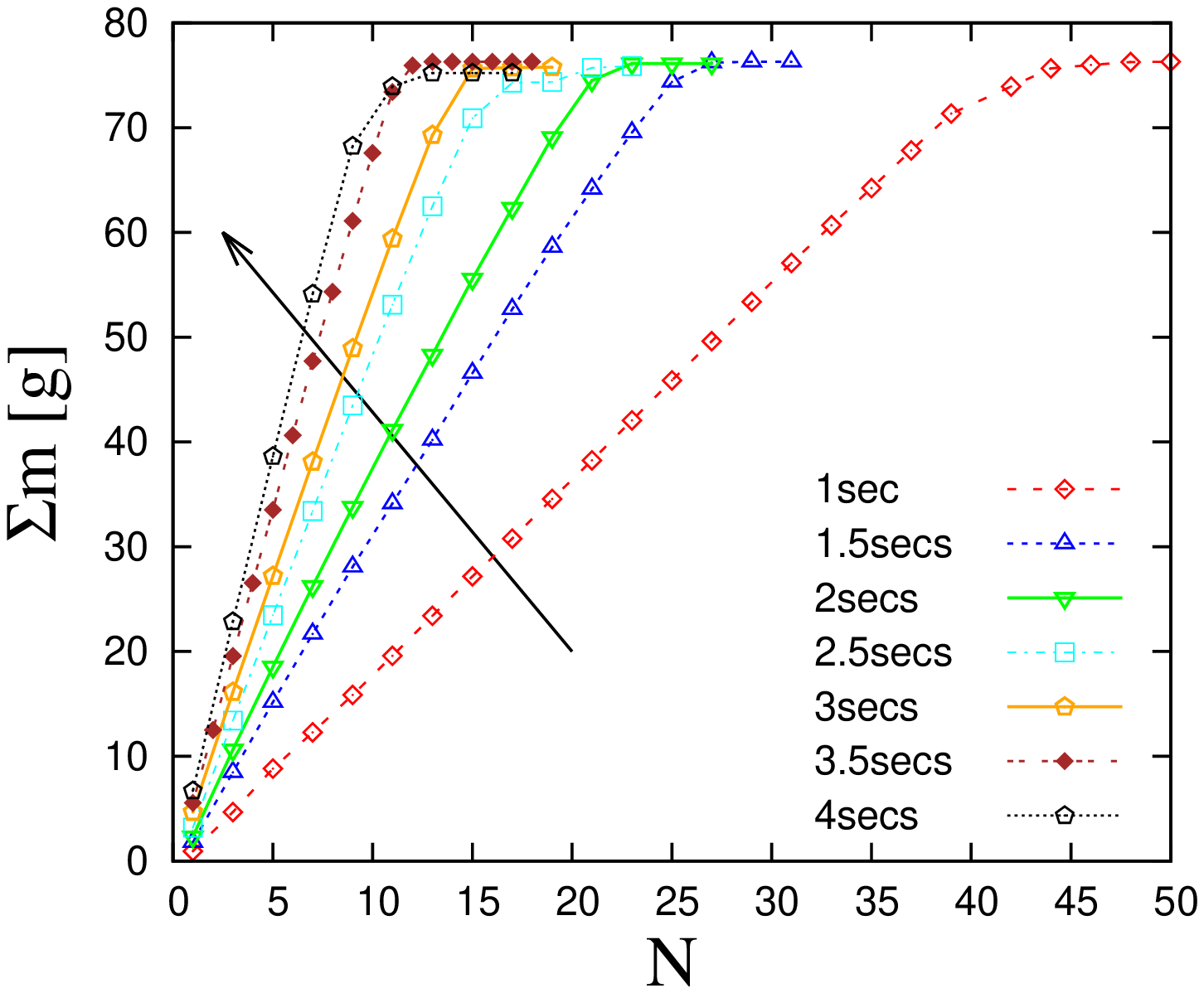}\label{timeres}}
\subfigure[]{\includegraphics[scale=0.28, angle=270]{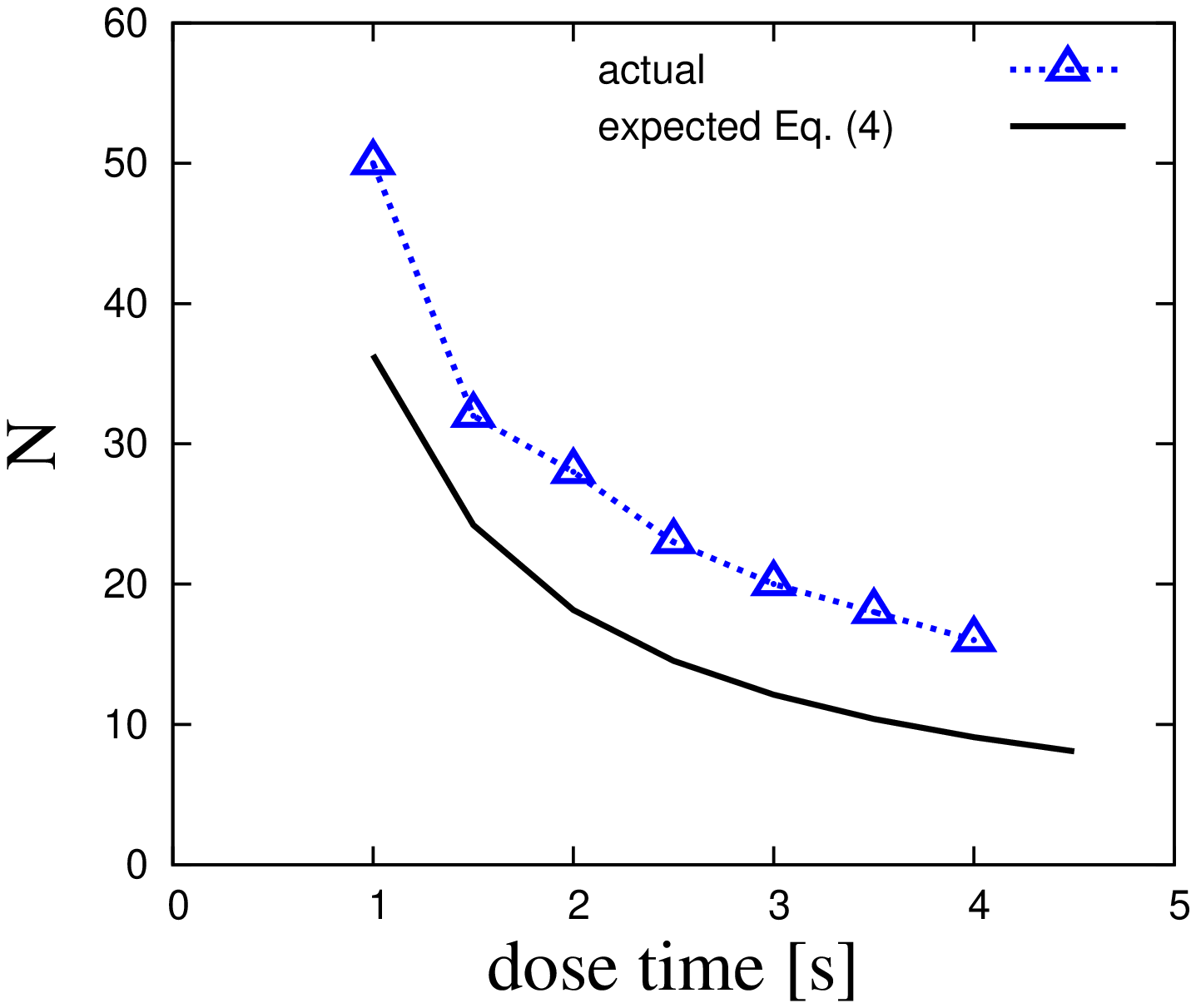}\label{doseno}}
\subfigure[]{\includegraphics[scale=0.28,angle=270]{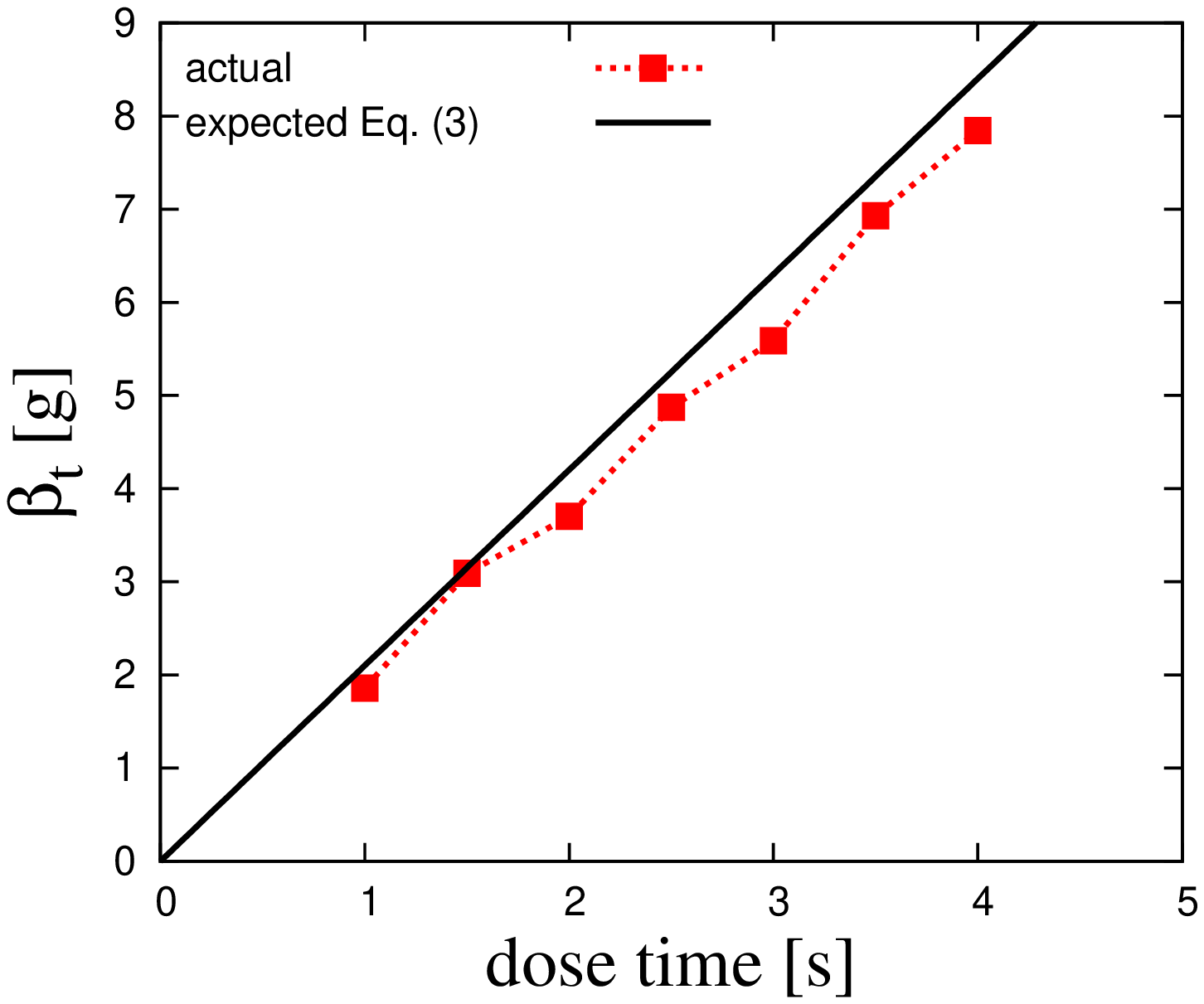}\label{doseslope}}\\
\caption{(a) Cumulative dosed mass from experiments plotted as function of the number of doses for different dose time, (b) number of doses recorded for the respective dose times, and
(c) mass per dose plotted for different dosage times. The solid black lines represent the expected mass $\mdose$ using Eq.\ (\ref{eq:mdoseexp}) and the expected doses $\ndose$ using Eq.\ (\ref{eq:ndoseexp}), 
as prediction. } %Arrows pointing right
\label{dosetimeparams}
\end{figure*}

To understand the effect of dosage time, in Fig.\ \ref{timeres}, we vary the dosage time from 1-4 s while keeping the initial powder mass in the canister and the rotation speed constant at 80g and 90rpm respectively. 
A first observation is the higher slope for longer dosing time, that leads to a decrease in the number of doses recorded. This is explained by the increased number of complete screw rotations as the dosage time is increased, thus allowing for an
increased mass throughput. In Fig.\ \ref{doseno}, we plot the number of doses for different dose time, and observe inverse proportionality. Also, the expected number of doses predicted using 
Eq.\ (\ref{eq:ndoseexp}) is lower. The decrease in the number of doses is faster between $t$= 1--1.5 s and then slows down as 
the time increases until $t=$4 s. In Fig.\ \ref{doseslope}, we plot the actual (red squares) and predicted  (solid black line) 
mass per dose $\bt$  taken from the cumulative dosed mass before saturation, for different dose time.
The predicted mass per dose is obtained using Eq.\ (\ref{eq:mdoseexp}) for an initial bulk density $\rhob \approx  4.71 \times 10^{-4} $ g/mm$^3$, coil pitch $p = $ 8.75 mm and coil radius 10.4 mm.  
We observe a linear increase in the mass per dose with increasing dosage time. The experimental mass per dose for the different dosage times is close to the predicted values with the predicted mass slightly 
higher. This indicates that less mass is being transported per dose, which could be due to the uneven, inhomogeneous re-filling of the coil during the dosing process and due to the small volume of the coil
that is not considered in  Eq.\ (\ref{eq:mdoseexp}).

\section{Numerical Results}
\label{sec:numresults}

In this section, we discuss the results from discrete element simulations of the dosing test. First, we compare the snapshots
of the particle bed surface with images taken from experiments. Next, we describe the process of calibration 
of the material parameters used in our simulations. As studied in the experiments, we show results 
on varying the dosage time, coil rotation speed and number of coils. Finally, we report on the macroscopic velocity
and density fields during the dosing process.

\subsection{Surface profile of the dosed material}
\label{sec:profile}

\begin{figure}[bt]
\subfigure[]{{\includegraphics[scale=0.29, width=3.85cm]{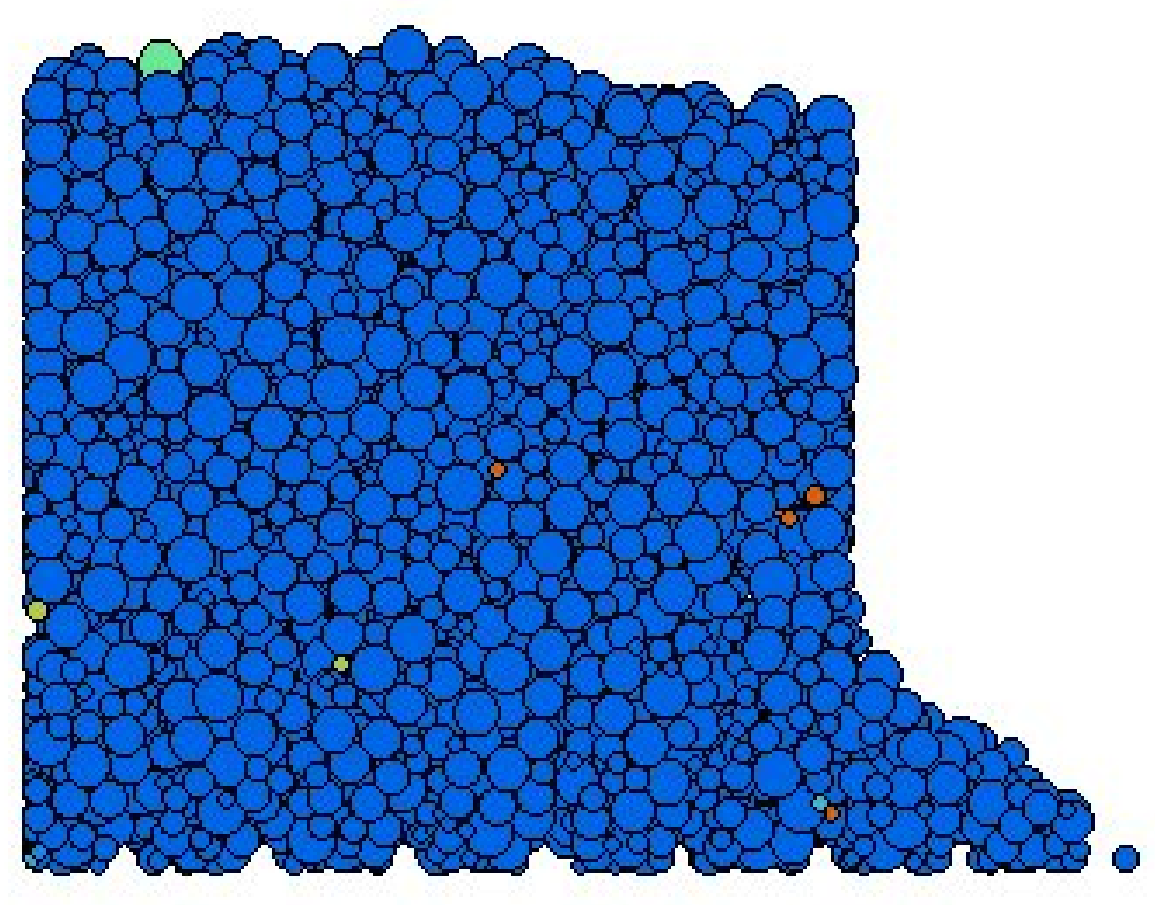}}}\hspace{-.18in}%\hspace{0.005em}
\setlength{\unitlength}{\columnwidth}
\!\!\!\!\!\!\!\!{\begin{picture}(0,0)
\put(-0.205,0.020){\vector(1,0){.220}}
\put(-0.206,0.019){\vector(0,1){.23}}
\put(-0.025,0.09){\line(0,1){.135}}  %.075
% \put(-.16,0.24){$t=0.2$}  %.-25,-.03
\end{picture}}
\label{figBetaa}
%%%%%%%%%%%%%%%%%%%%%%%%%%%%%%
\subfigure[]{{\includegraphics[scale=0.29, width=4.1cm]{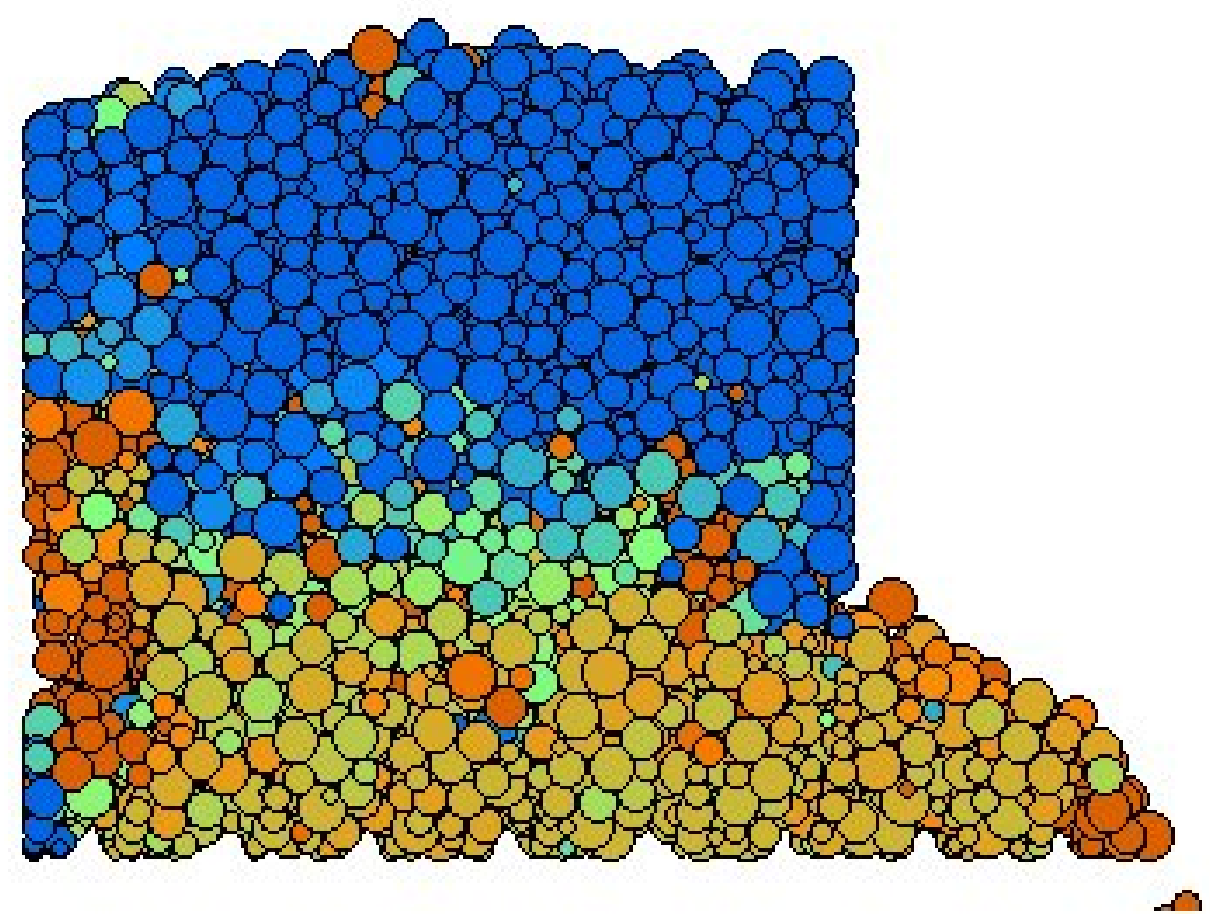}}}\hspace{-.18in}%\hspace{0.005em}
\setlength{\unitlength}{\columnwidth}
\!\!\!\!\!\!\!\!{\begin{picture}(0,0)
\put(-0.225,0.020){\vector(1,0){.220}}
\put(-0.223,0.020){\vector(0,1){.23}}
\put(-0.032,0.09){\line(0,1){.135}}  %.075
% \put(-.16,0.24){$t=0.5$}  %.-25,-.03
\end{picture}}
\label{figBetab}
% %%%%%%%%%%%%%%%%%%%%%%%%%%%
\subfigure[]{{\includegraphics[scale=0.30, width=4.1cm]{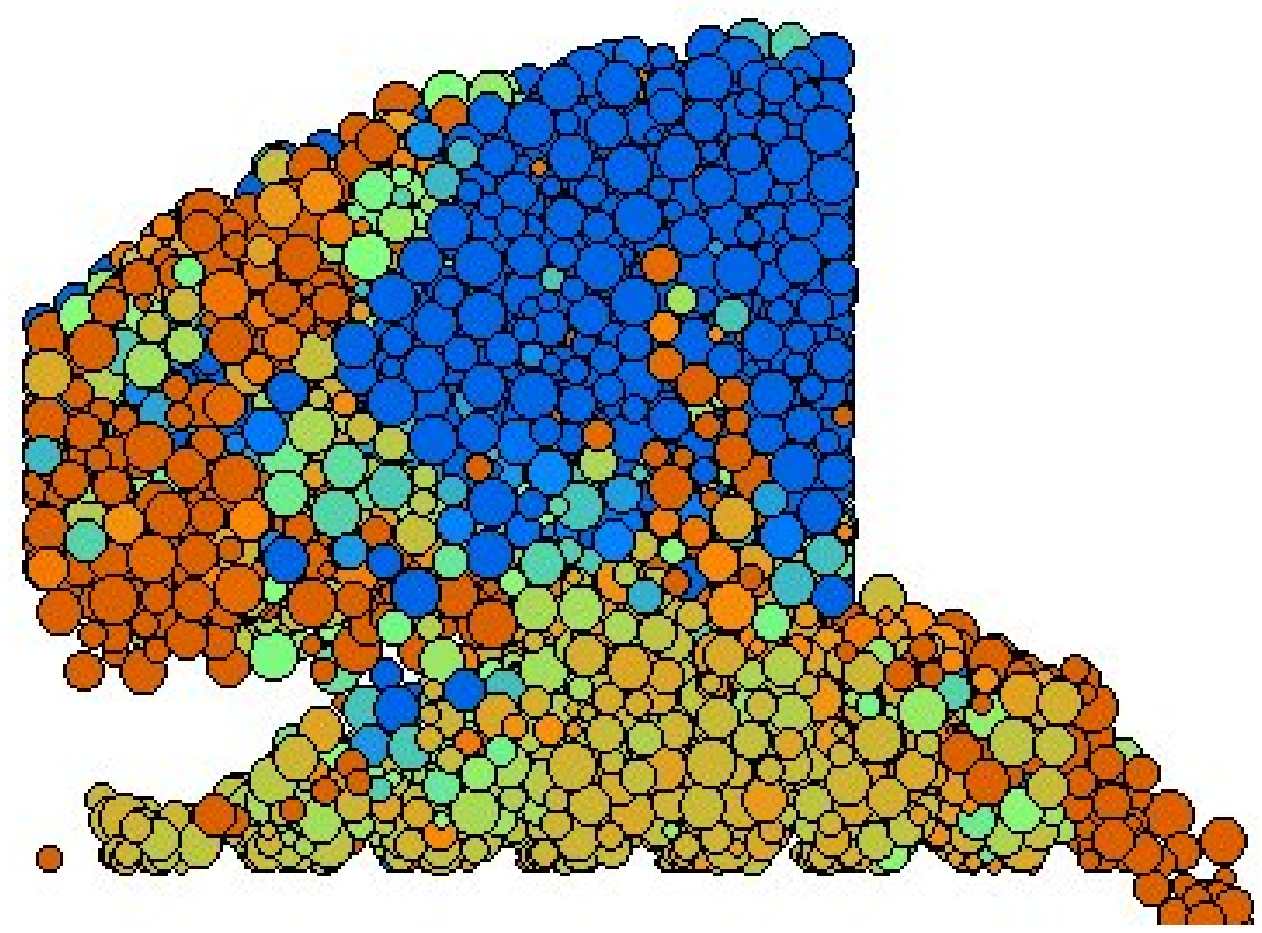}}}\hspace{-.18in}%\hspace{0.005em}
\setlength{\unitlength}{\columnwidth}
\!\!\!\!\!\!\!\!{\begin{picture}(0,0)
\put(-0.225,0.020){\vector(1,0){.220}}
\put(-0.223,0.020){\vector(0,1){.23}}
\put(-0.031,0.09){\line(0,1){.135}}  %.075
% \put(-.16,0.24){$t=0.7$}  %.-25,-.03
\end{picture}}
\label{figBetac}
% %%%%%%%%%%%%%%%%%%%%%%%%%%%%%%%%
\subfigure[]{{\includegraphics[scale=0.29, width=4.1cm]{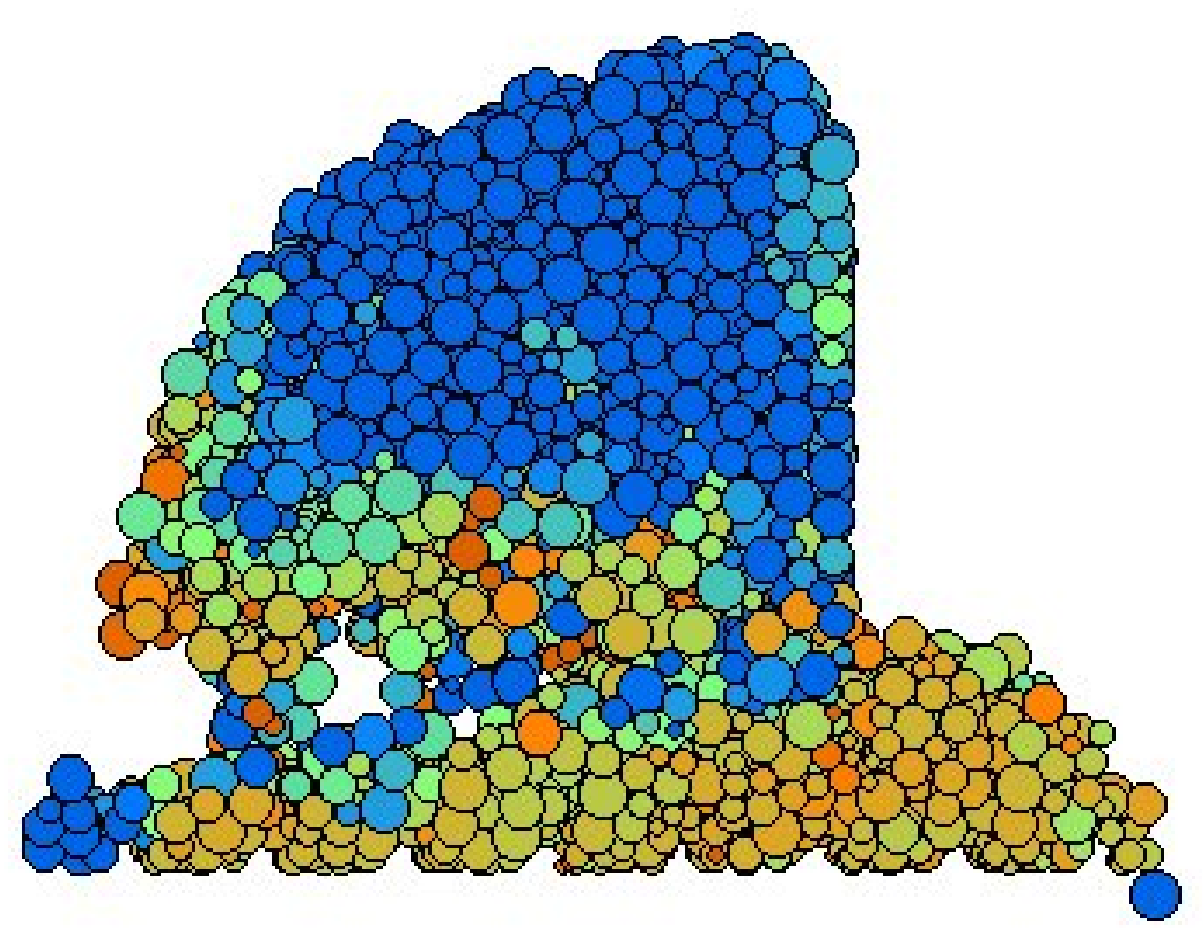}}}\hspace{-.18in}
\setlength{\unitlength}{\columnwidth}
\!\!\!\!\!\!\!\!{\begin{picture}(0,0)
\put(-0.232,0.020){\vector(1,0){.220}}
\put(-0.232,0.020){\vector(0,1){.23}}
\put(-0.040,0.09){\line(0,1){.135}}  %.075
% \put(-.16,0.24){$t=0.9$}  %.-25,-.03
\end{picture}}
\label{figBetad}\\
% %%%%%%%%%%%%%%%%%%%%%%%%%%%%%%%%
\subfigure[]{{\includegraphics[scale=0.29,width=2.9cm]{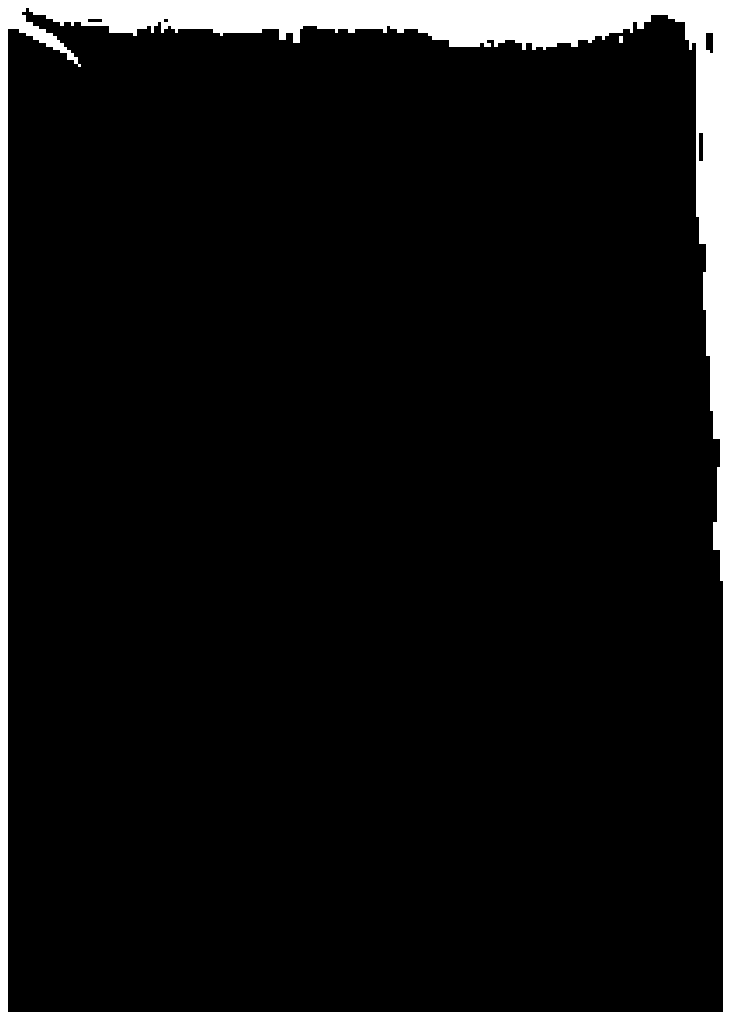}}}\hspace{0.2in}
\setlength{\unitlength}{\columnwidth}
\!\!\!\!\!\!\!\!{\begin{picture}(0,0)
% \put(-0.225,0.020){\vector(1,0){.235}}
% \put(-0.225,0.020){\vector(0,1){.23}}
% \put(-0.041,0.09){\vector(0,1){.16}}  %.075
% \put(-.16,0.24){$t=0.2$}  %.-25,-.03
\end{picture}}
\label{figBetae}
% %%%%%%%%%%%%%%%%%%%%%%%%%%%%%%%%
\subfigure[]{{\includegraphics[scale=0.29,width=2.9cm]{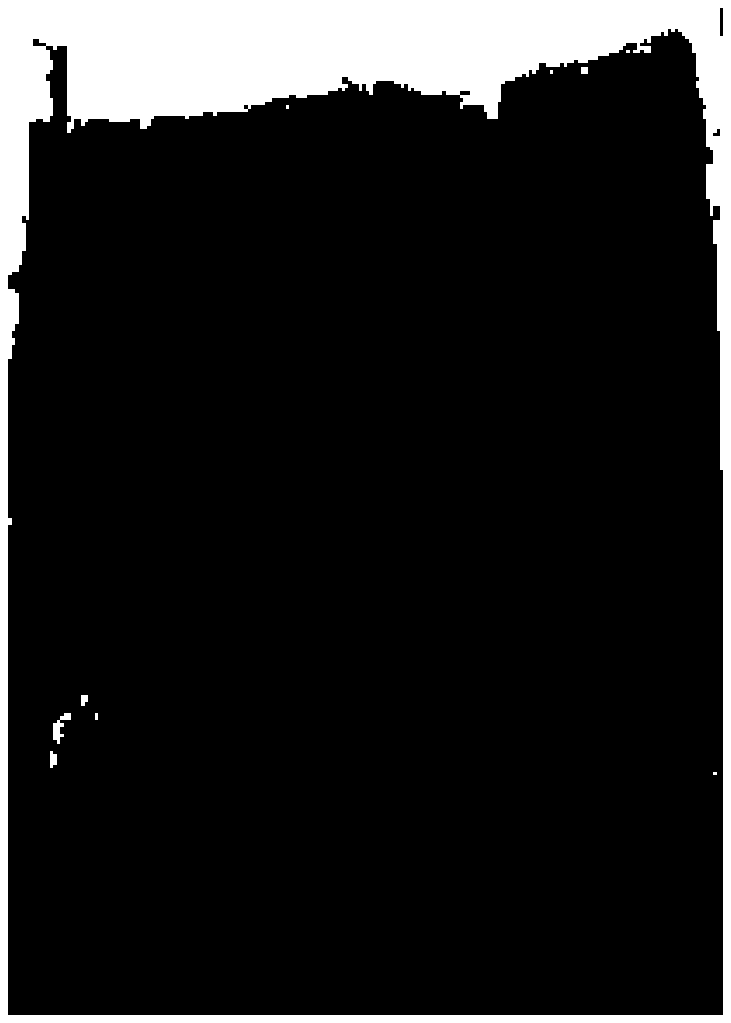}}}\hspace{0.2in}
\setlength{\unitlength}{\columnwidth}
\!\!\!\!\!\!\!\!{\begin{picture}(0,0)
% \put(-0.225,0.020){\vector(1,0){.235}}
% \put(-0.225,0.020){\vector(0,1){.23}}
% \put(-0.041,0.09){\vector(0,1){.16}}  %.075
% \put(-.16,0.24){$t=0.2$}  %.-25,-.03
\end{picture}}
\label{figBetaf}
% %%%%%%%%%%%%%%%%%%%%%%%%%%%%%%%%
\subfigure[]{{\includegraphics[scale=0.29,width=2.9cm]{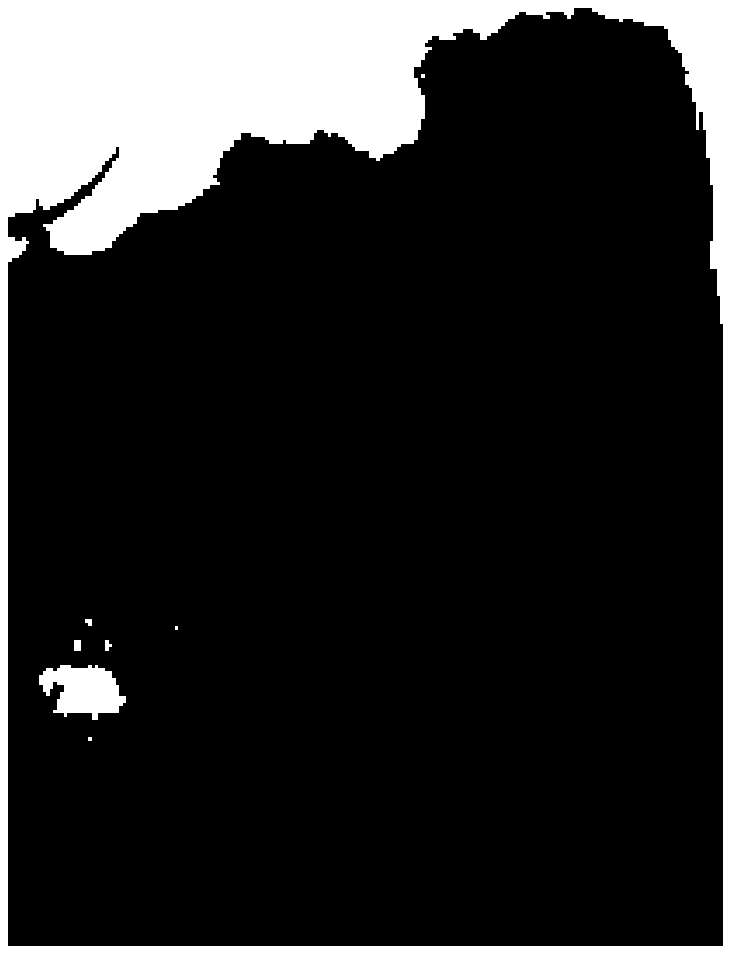}}}\hspace{0.2in}
\setlength{\unitlength}{\columnwidth}
\!\!\!\!\!\!\!\!{\begin{picture}(0,0)
% \put(-0.225,0.020){\vector(1,0){.235}}
% \put(-0.225,0.020){\vector(0,1){.23}}
% \put(-0.041,0.09){\vector(0,1){.16}}  %.075
% \put(-.16,0.24){$t=0.2$}  %.-25,-.03
\end{picture}}
\label{figBetag}
% %%%%%%%%%%%%%%%%%%%%%%%%%%%%%%%%
\subfigure[]{{\includegraphics[scale=0.29,width=2.9cm]{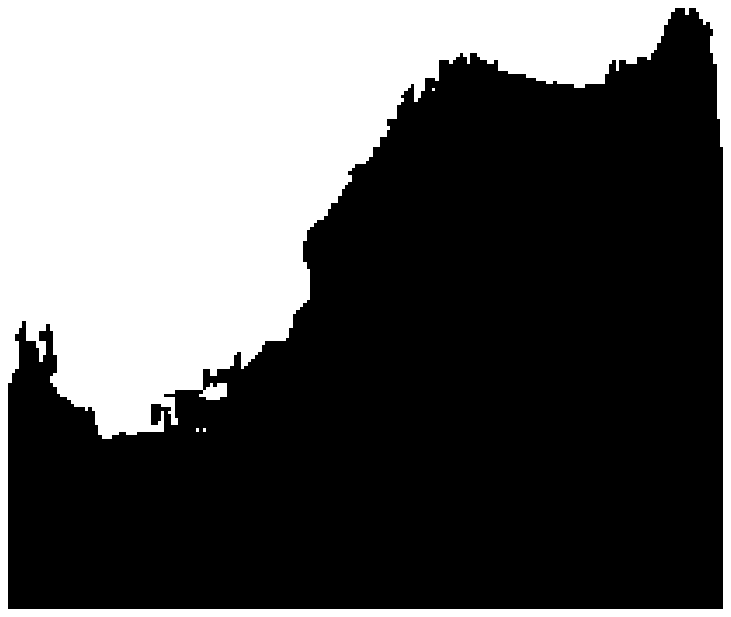}}}%\hspace{12in}
\setlength{\unitlength}{\columnwidth}
\!\!\!\!\!\!\!\!{\begin{picture}(0,0)
% \put(-0.225,0.020){\vector(1,0){.235}}
% \put(-0.225,0.020){\vector(0,1){.23}}
% \put(-0.041,0.09){\vector(0,1){.16}}  %.075
% \put(-.16,0.24){$t=0.2$}  %.-25,-.03
\end{picture}}
\label{figBetah}
% %%%%%%%%%%%%%%%%%%%%%%%%%%%%%%%%

\caption{Snapshot of the time evolution of the simulation during the dosing test with time increasing from (a--d) and (e-h), 
respectively. (a--d) are taken from simulation while 
comparable snapshots (e--h) are image processed experimental visualizations of the powder profile. Colors/shades in (a--d)
indicate the kinetic energy of the particles with blue (static) and orange (dynamic) particles. For the simulation, parameters are 
$\bkc=0.872$ and $\mu=0.5$. The coil is not shown for clarity. \vspace{-.18in}}
\label{figBeta}

\end{figure}

As a first step to gain insights into the dosing process,  we show exemplary snapshots of the time evolution of the 
surface profile of bulk sample during a typical simulation in Figs.\ \ref{figBeta}(a-d). For this study, the initial mass
in the box is set at 60grams while the coil with the narrow pitch (8 complete turns) is used. Fig.\ \ref{figBeta}(a) shows 
the state of the bulk sample sample after the first 2 seconds where the particles have been allowed to settle. At this point, the
kinetic energy of the particles are close to zero since they are non-mobile. As the coil begins to turn in Fig.\ \ref{figBeta}(b) after the relaxation phase,
particles within the area of the coil begin to move leading to an increase in their kinetic energy, as seen from the bright colors in the lower part 
of the box. In general, particles around the uppermost layer of the box remain largely static while the the 
region where the kinetic energy is highest can be
seen around the rear end of the coil. Moving further in time to Fig.\ \ref{figBeta}(c), we find that the emptying of 
the box occurs faster at the rear (left) end of the
box, thereby causing avalanches as the void left due to the emptying of the box is filled. In addition to this, we observe in 
some cases, arches forming above the coil, where the 
void created below the screw is visible. We must also point out that particles closest to the right wall of the box remain static 
and they only collapse into the coil at the base
as an increased amount of powder is dispensed from the box as shown in Fig.\ \ref{figBeta}(d). %The collapse of the bulk particles closest to the right wall of the box continues  

Along with this, in Figs.\ \ref{figBeta}(e-h) we show image processed visualizations of the experimental powder profile during the 
dosing process. From the initial solid, bulk powder in
 Fig.\ \ref{figBeta}(e), we observe a progressive change in the powder surface profile with the canister emptying faster from its 
left rear end. Arches forming on the lower left side of the
box above the coil is also seen leading to avalanches and collapse of the powder around this region.

In summary, comparing the experimental and simulation profiles of the powder surface, we observe that the essential features observed in the experiment, 
namely the faster emptying at the rear end of the coil and arches forming during ongoing dosage are reproduced in the simulation. Also, we
must point out that the faster emptying at the rear end of the 
coil is due to the design of the coil which can be mitigated through the use of conical inserts in the coil \cite{ramaioli2007granular}. 
In the next sections, we will focus on a 
quantitative comparison between experiments and simulation.

\subsection{Calibration and Sensitivity Studies}
\label{sec:calib}

The particles used in the simulation can be seen as meso-particles consisting of an agglomerate of 
other smaller particles. Due to this, it is important that their material properties are carefully selected 
based on sensitivity studies of how each parameter influence the dosing process in comparison to the experiment. 

In order to obtain relevant parameters unique for our problem, we perform  various studies
in order to test the sensitivity of the essential material parameters, namely interparticle friction and cohesion
during the dosing process. To achieve this, several simulations were run where the interparticle friction
is fixed in each case and cohesion is varied. Note that for each simulation, we obtain data on the 
cumulative dosed mass and the number of doses. From each simulation, the respective mass per dose $\beta$
are obtained within the linear region where initial conditions and other artefacts due to arching are
absent. The mass per dose $\beta$ is then systematically compared for different interparticle friction and cohesion and 
bench-marked against the obtained experimental $\beta$ value. We choose $\beta$  as a calibration parameter 
since it is largely independent of the initial mass (see Fig.\ \ref{massres}). The For the sake of brevity, this calibration
procedure is performed on using a total mass of 48grams in the box and the narrow pitch coil with 8 complete turns. We attempted
a calibration with higher masses as compared with the experiments but we observe that due to arching occurring when cohesion is 
high, the plot of the 
cumulative dosed mass becomes non-linear. This made defining an appropriate $\beta$ challenging therefore requires further work.
In the mean time, we focus the calibration with the lower mass.

%%%%%%%%%%%%%%%%%%%%

\begin{figure} [!ht]
\centering
\includegraphics[scale=0.60, angle=270]{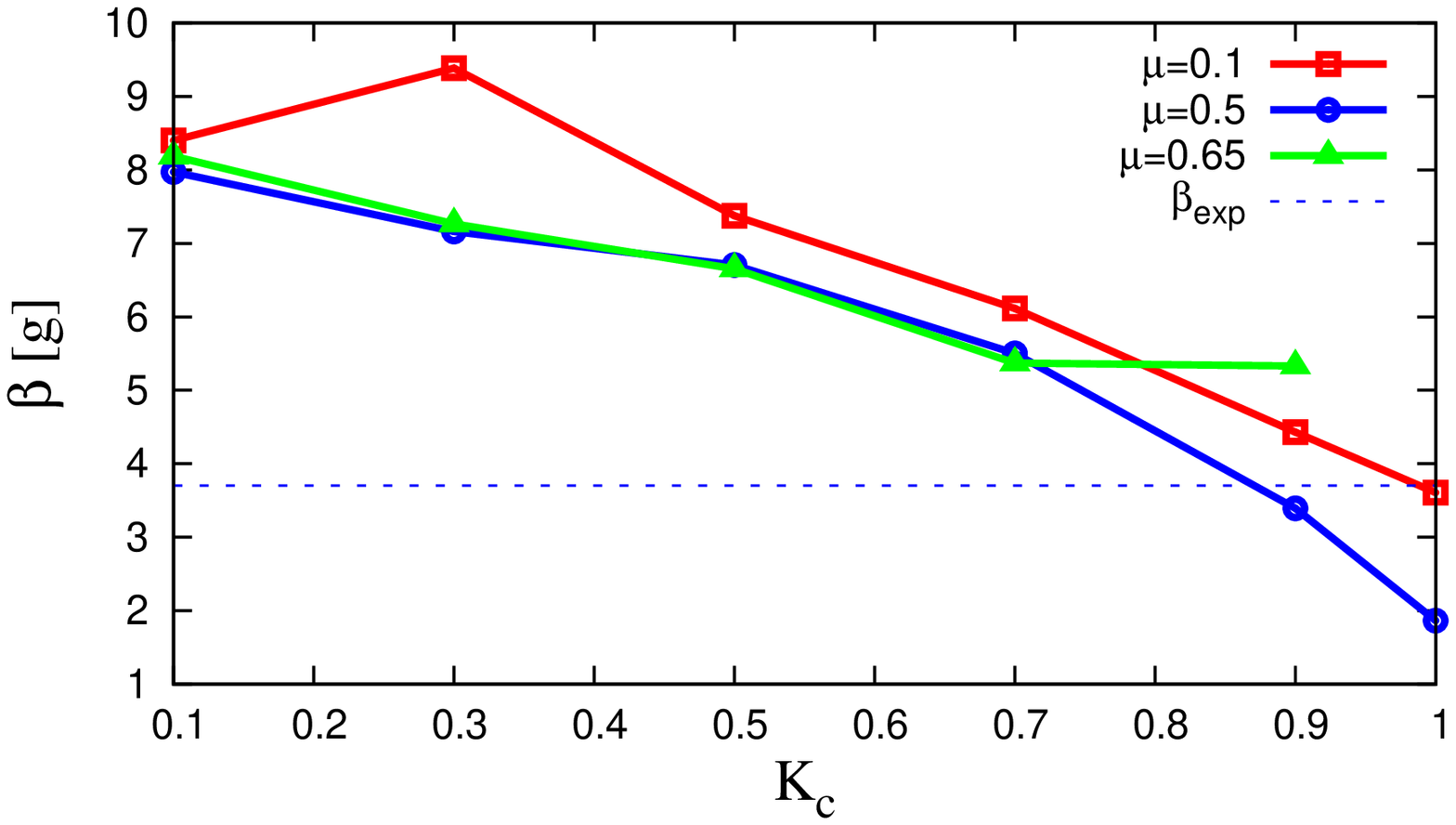}
\caption{Calibration of the cohesive stiffness $\bkc= \kc/k$ and inter particle friction $\mu$. We plot the mass per dose $\beta$ for different $\bkc$ and different $\mu$ as
given in the inset. The dotted horizontal line shows the experimental $\beta$ value.}
\label{kc48}
\end{figure}

In Fig.\ \ref{kc48}, we show the mass per dose $\beta$, plotted against the interparticle cohesive stiffness $\bkc$ and different 
interparticle friction coefficient $\mu$. The horizontal dotted line shows the mass per dose obtained in the experiment with 
value 3.702g/dose. A first observation is the consistent decrease of $\beta$ with increasing $\bkc$ for all friction. This is due to 
reduced flowability of the bulk sample with increasing cohesion. We note however that for the highest friction, we observe
a slight increase in the $\beta$ values obtained at high cohesion. This is a consequence of arching that sets in due to high cohesion
causing a bridge in the flow especially in the region above the coil. This leads to highly unsteady mass throughput 
from the box.

Comparing the data for 
different friction, we observe a decrease in $\beta$ with increasing $\mu$. Increased interparticle friction leads to an
an increased resistance to flow which reduces the rate at which the material is being dispensed out of the box and consequently
lower $\beta$. Similar to what is found in other studies, for interparticle friction within the range $\mu=$0.5 and 0.65,
the effect becomes less strong as seen in the saturation and collapse of $\beta$. % for the configurations with the highest friction between $\mu =$0.5 and 0.65.

As seen from Fig.\ \ref{kc48}, the experimental measured mass per dose (dotted horizontal line) intersects with the different friction data at different 
points leading to different possible $\bkc$ values. %as shown in {\color{red} Table}. 
A choice therefore has to be made of the appropriate
$\bkc$ which reproduces the experiments and leads to the least variability between successive doses in the simulations. In this case, we choose the lowest possible
$\bkc$ which gives the match with the experimental $\beta$ value at $\bkc=0.872$ and $\mu = 0.50$.

\subsection{Comparison with Experiments}
\label{sec:valide}

In order to test the validity of the interparticle friction and cohesion parameters obtained from the
calibration test, we perform simulation setting $\bkc=0.872$ and $\mu=0.5$. We then compare the simulation results with
experiments. For both experiment and simulation,
the narrow coil with 8 turns is used. For each dose, the coil is rotated at a speed of 90rpm for 2 seconds.

We observe that for the first few doses, the experimental and numerical dosed masses obtained are slightly
different -- with the simulation slightly under-predicting the experimental masses. This is possibly arising from the different initial preparation and the randomness of the initial states.
After the first few doses, the simulation is observed to compare well with experiments with both datasets collapsing on each other. By comparing the individual points on the 
cumulative dosed mass plots between experiment and simulation, we obtain a maximum variation in mass per dose of less than 9 percent. This is comparable to the variation of about 5 percent obtained for
experiments with different masses (see section\ \ref{sec:inimass}).

\subsection{Parametric Studies}
\label{sec:paramsstd}

In this subsection, we will discuss the numerical results of parametric studies on the dosing experiments.
Similar to the experiments, we investigate the effect of varying the dosage time and the number of coils.
Although not studied in the experiment, we also look at the effect of higher rotation speeds during the dosing
action.

\begin{figure*}[!ht]
 \centering
\subfigure[]{\includegraphics[scale=0.44, angle=270]{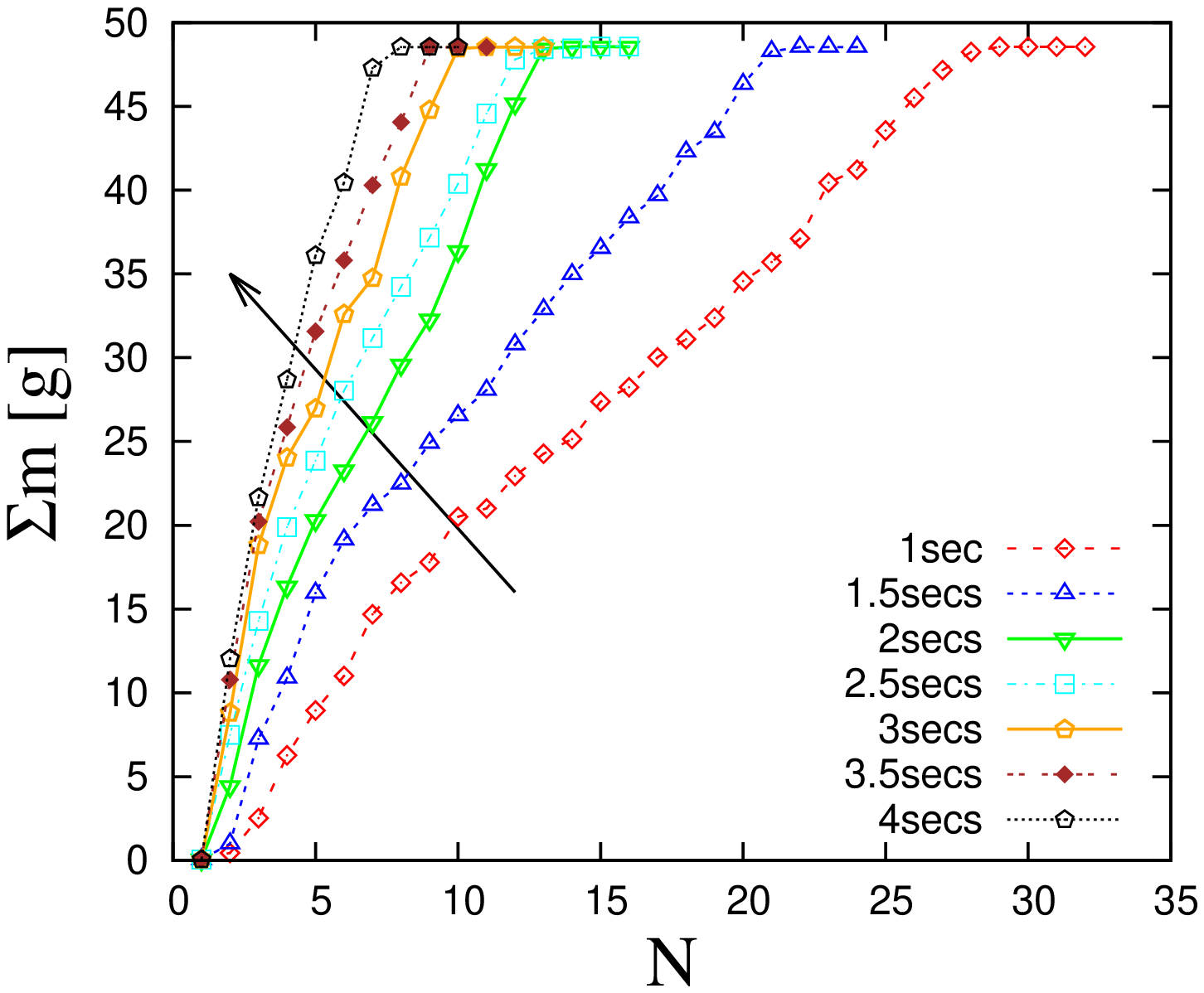}\label{timesimres}}
%\subfigure[]{\includegraphics[scale=0.29, angle=270]{dtime_doseno_sim.ps}\label{dosenosim}}
\subfigure[]{\includegraphics[scale=0.44,angle=270]{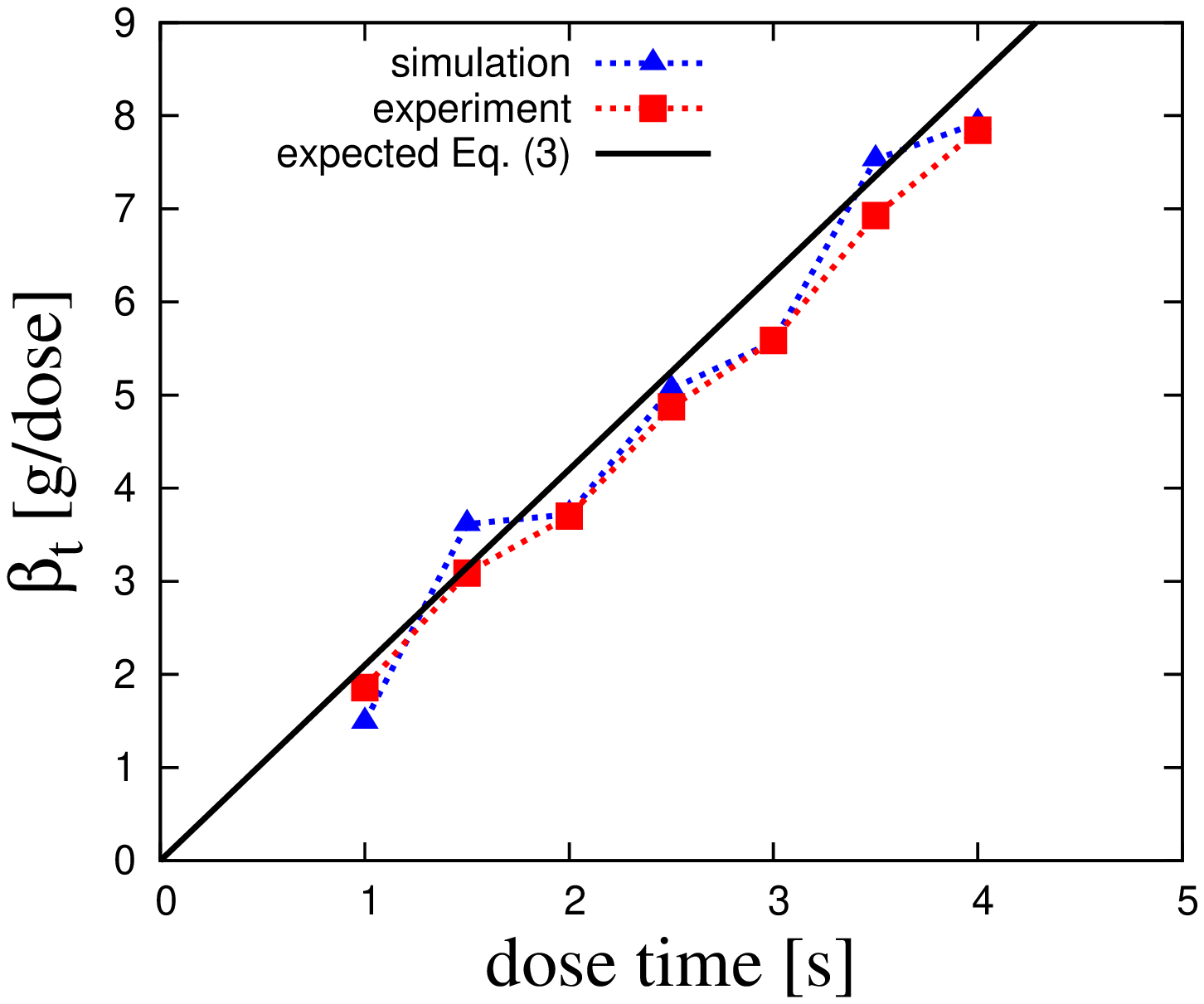}\label{doseslopesim}}\\
\caption{ (a) Cumulative dosed mass from simulation plotted as function of the number of dose for different dose time; 
(b) mass per dose obtained from simulation (from the first few doses) and experiments for different dose times. The solid black line represents the expected mass $\mdose$ using Eq.\ (\ref{eq:mdoseexp}) as prediction.} %Arrows pointing right
\label{simdosetimeparams}
\end{figure*}

In Fig.\ \ref{simdosetimeparams}, we plot the cumulative dosed mass as function of the number of dose 
for different dosage times. The initial mass in the canister is 48grams while the interparticle friction
and cohesion are kept constant at 0.5 and 0.872, respectively. From Fig.\ \ref{timesimres}, we observe
that the cumulative dosed mass increases slightly non-linearly as the number of doses increases. The effect of slight arching and inhomogeneous density is evident by the slight reduction in the mass per dose
as the number of dose increases. Due to this, the mass per dose is obtained over the first few doses before arching sets in. The number of dose is obtained when the cumulative dosed mass 
does not change for three consecutive doses. 

%%%%%%%%%%%%%%%%%%%%

The mass per dose $\beta_t$ for different dosage time is compared between simulation and experiment in Fig.\ \ref{doseslopesim}. For all simulations, the mass per dose is obtained from the first few doses 
as the slope of the cumulative dosed mass in the linear region where the cumulative dosed mass is less than 15 grams.
 Recall that the calibration was done at a dose time of 
2 seconds while parameters obtained are then used for the other dose times. The mass per dose is found to increase linearly with the dose time in simulation and experiments. 
The mass per dose obtained from experiments and simulation for different dosage times are slightly lower than the prediction from Eq.\ (\ref{eq:mdoseexp}).

\begin{figure*}[!ht]
 \centering
\subfigure[]{\includegraphics[scale=0.29,angle=270]{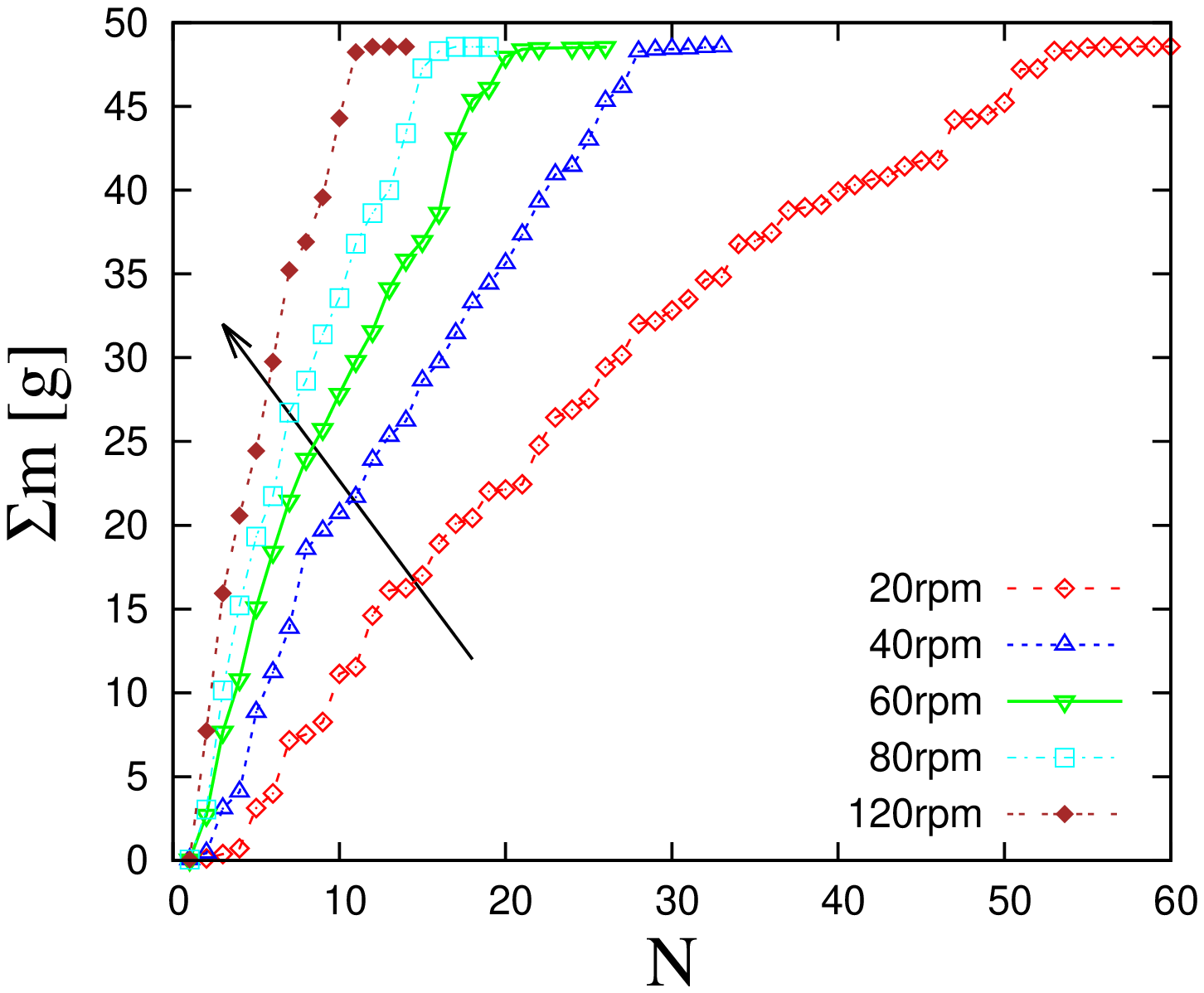}\label{simrotspeed}}
\subfigure[]{\includegraphics[scale=0.29,angle=270]{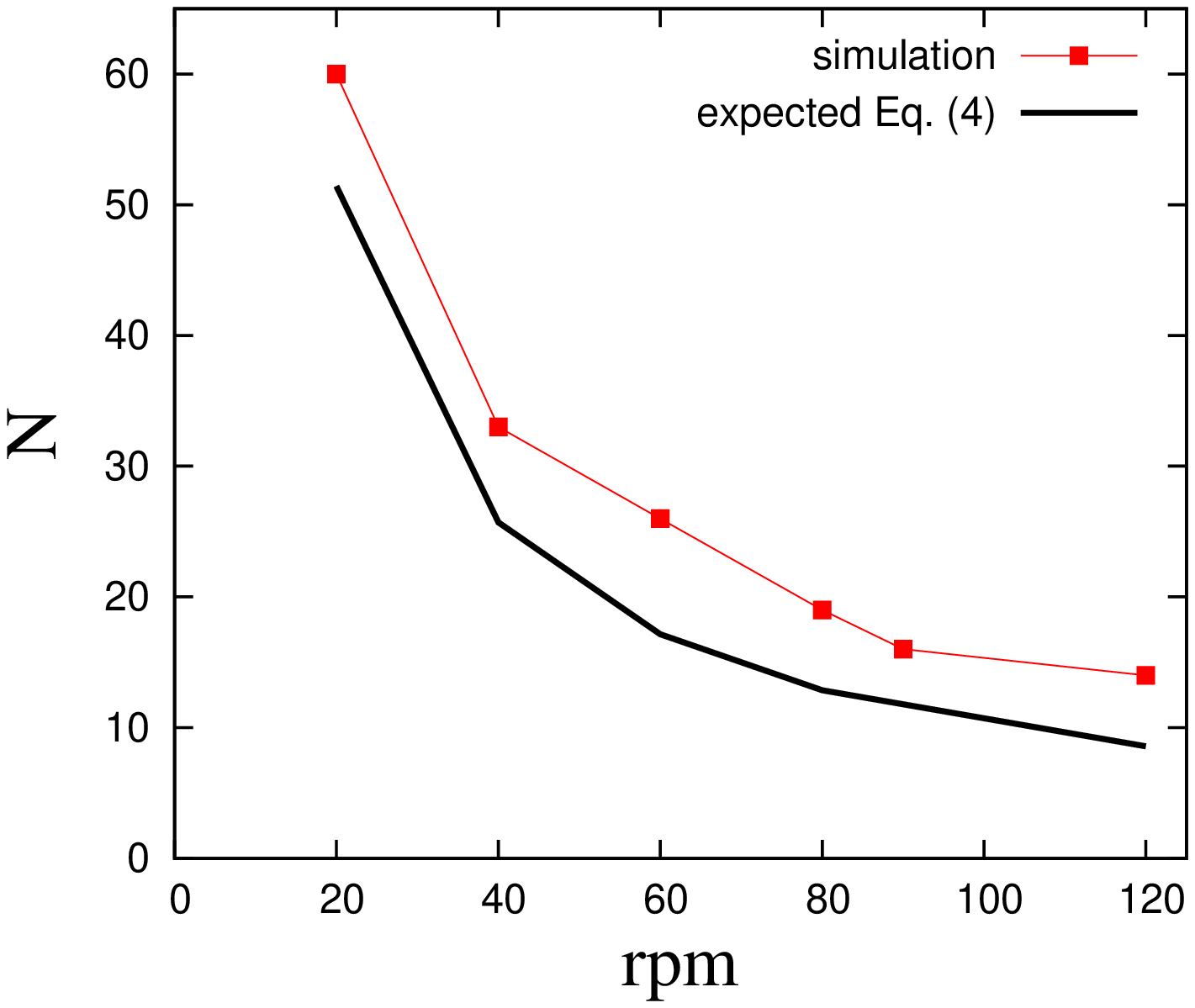}\label{rotspeeddoseno}}
\subfigure[]{\includegraphics[scale=0.29,angle=270]{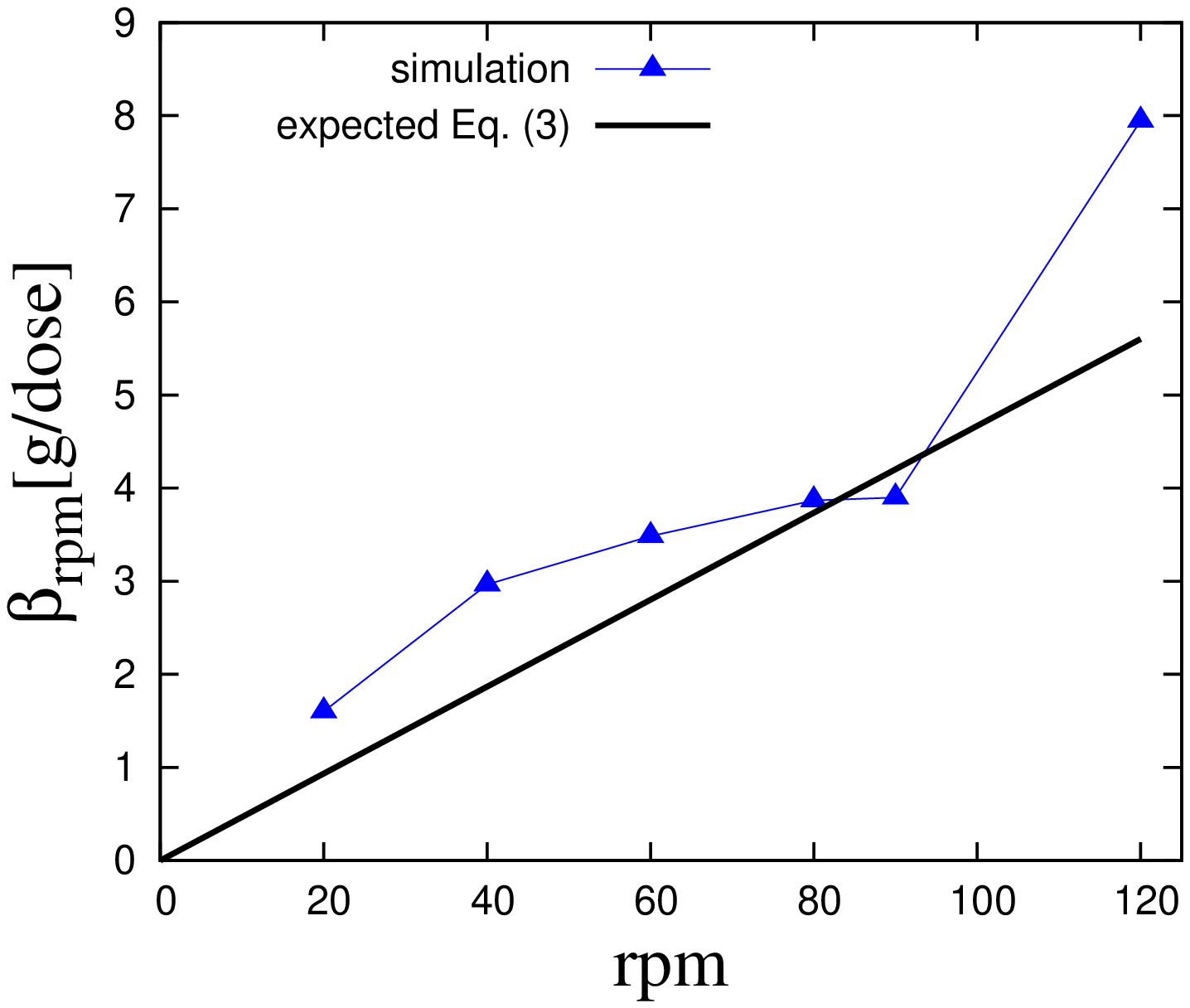}\label{rotspeedslope}}\\
\caption{(a) Cumulative dosed mass from simulations plotted as function of the number of doses for different rotation speeds
(b) Number of dose and (c) mass per dose $\beta_{rpm}$ obtained from the respective simulations for different rotation speeds.
The solid black lines represent the expected mass $\mdose$ using Eq.\ (\ref{eq:mdoseexp}), and the expected doses $\ndose$ using Eq.\ (\ref{eq:ndoseexp}), as prediction.}
\label{rotspeedparams}
\end{figure*}

%%%%%%%%%%%%%%%%%%%%%%%%

The effect of varying the rotation speed is presented in Fig.\ \ref{rotspeedparams}. 
For these simulations, the dose time is fixed to 2 seconds with while the rotation speed is varied.
It is evident that the box empties faster with increasing
rotation speed. Also, the number of doses required for the complete emptying of the box, shown in Fig.\  \ref{rotspeeddoseno}, 
is found to decrease fast between 20rpm and 40rpm followed by a much slower decrease upon further increase in the 
coil rotation speed. The slope (mass per dose) of the cumulative dosed mass, obtained from the first few doses for different rotation speeds $\beta_{rpm}$, shown
in Fig.\ \ref{rotspeedslope}, increases with rotation speed -- similar to $\beta_t$ observed 
when the dosage time is varied. Here, interestingly the expected mass per dose, Eq.\ (\ref{eq:mdoseexp}) mostly under-predicting the simulation, 
possibly due to on-going refilling of the coil during rotation and variation in the bulk density from compaction and avalanches occurring during the simulation.

\begin{figure} [!ht]
\centering
\subfigure[]{\includegraphics[scale=0.29,angle=270]{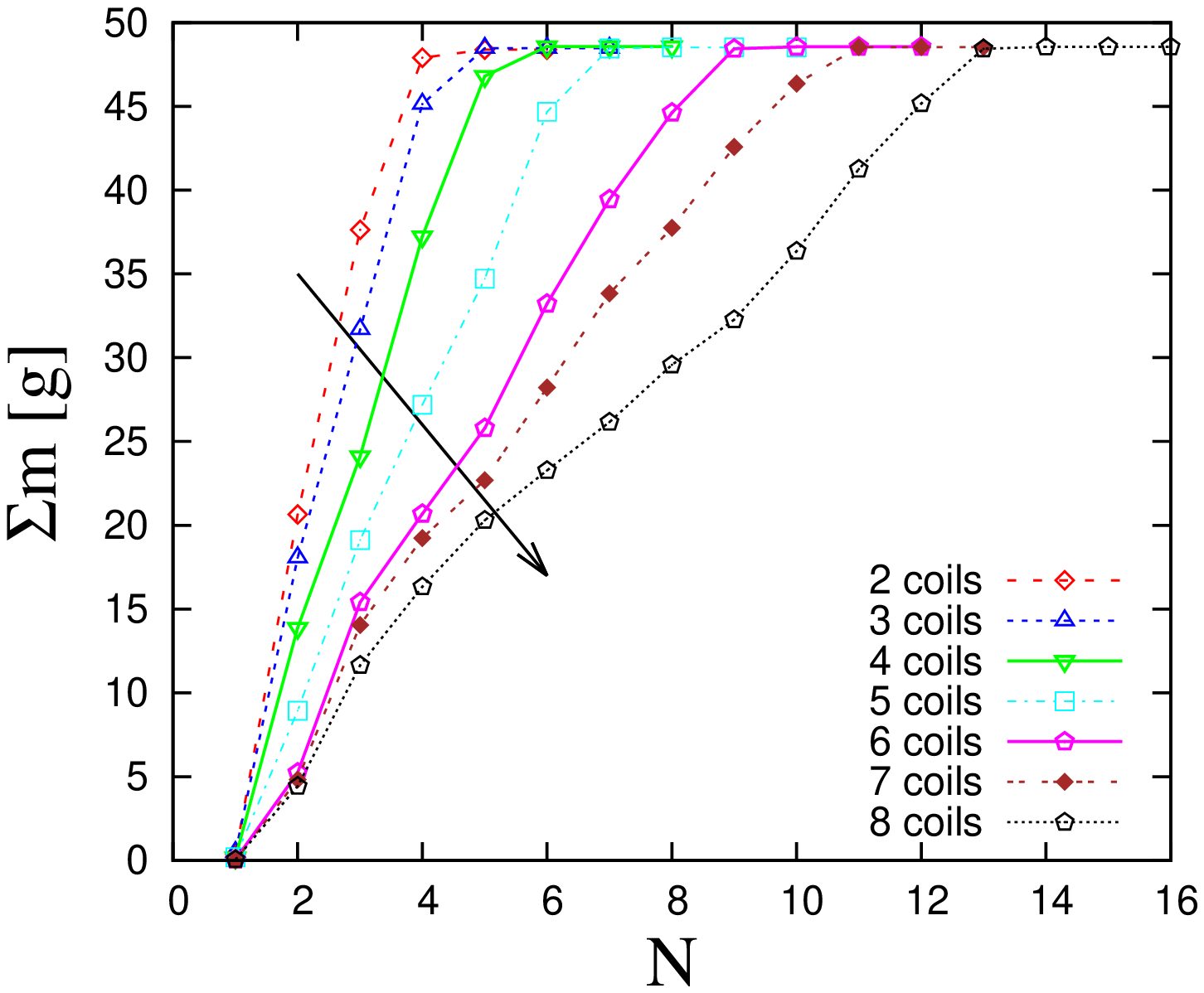}\label{simnocoils}}
\subfigure[]{\includegraphics[scale=0.29,angle=270]{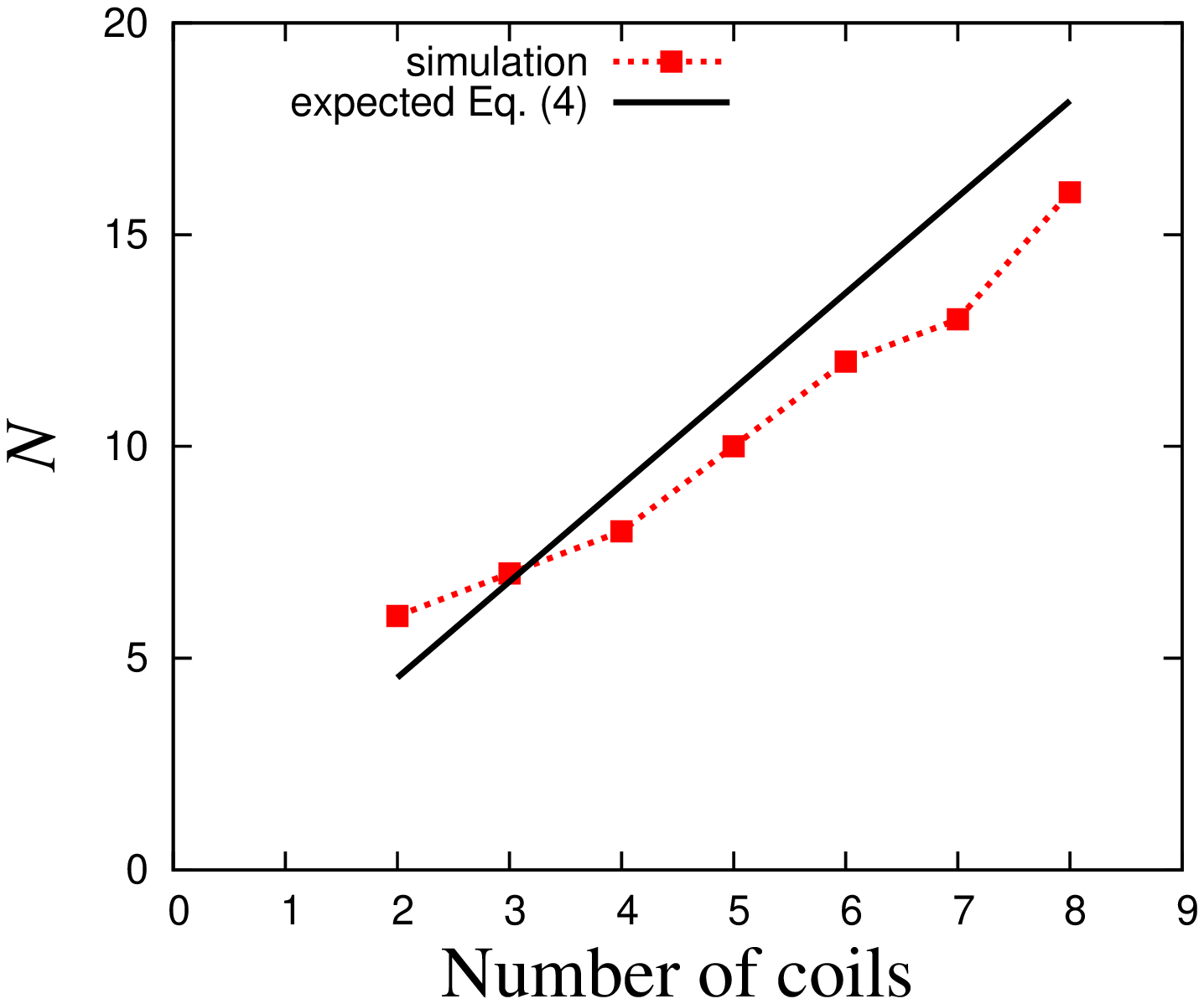}\label{coildoseno}}
\subfigure[]{\includegraphics[scale=0.29,angle=270]{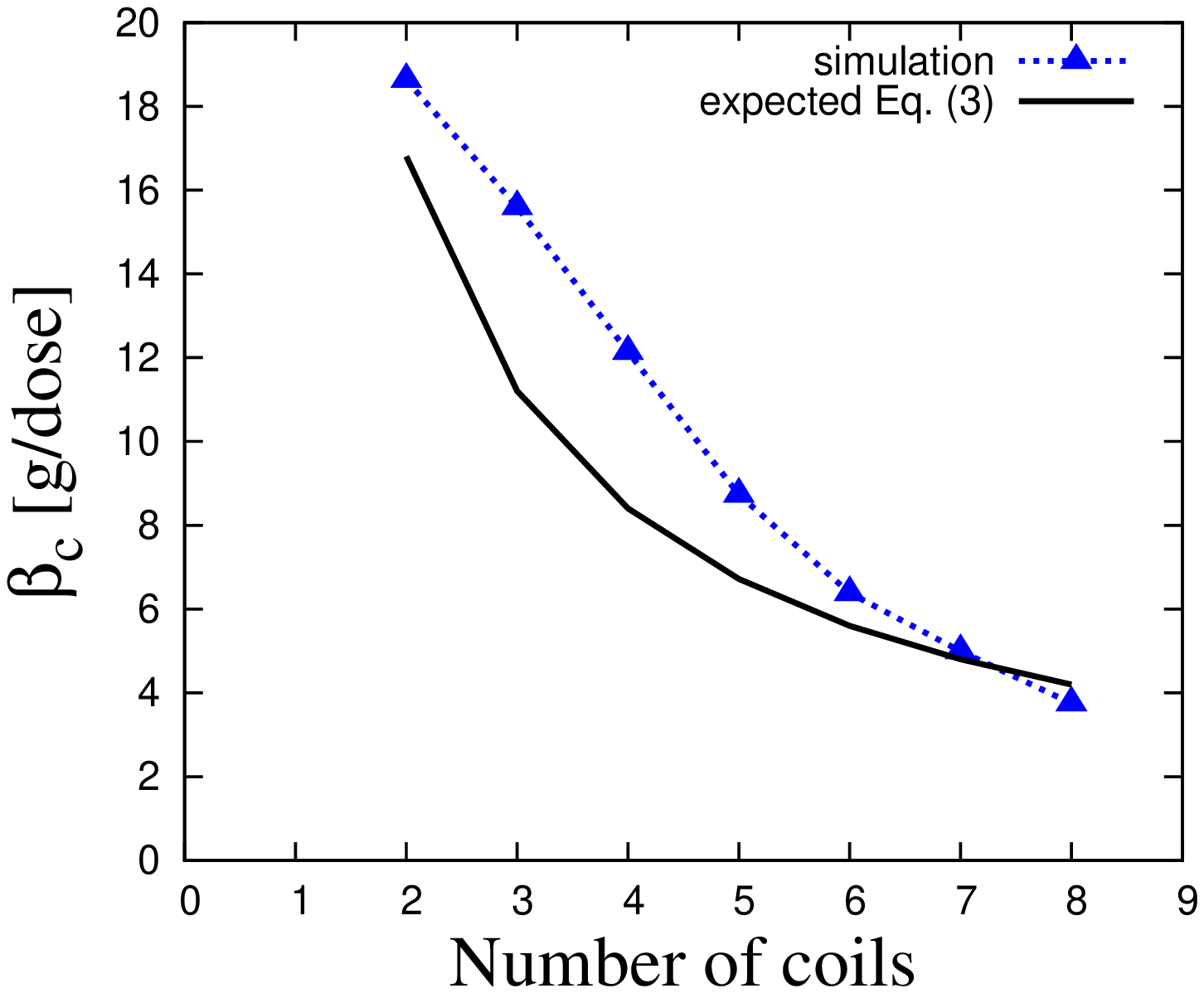}\label{coilslope}}
\caption{ (a) Cumulative dosed mass from simulation plotted as function of the number of doses for different number of  coils. The arrow indicate the 
decreasing trend with increasing number of coils. 
(b) Number of doses and (c) mass per dose $\beta_{c}$ obtained from the respective simulations for different number of coils.
The solid black lines represent the expected mass $\mdose$ using Eq.\ (\ref{eq:mdoseexp}), and the expected doses $\ndose$ using Eq.\ (\ref{eq:ndoseexp}), as prediction.}
\label{numberofcoils}
\end{figure}

Simulation results on the effect of varying the number of coils from 2 to 8 coils are shown in Fig.\ \ref{numberofcoils}. 
Increasing the number of coils essentially means reducing the pitch of the coil such that the simulation with two coils
has the widest pitch. An increase in the number of coils is accompanied by an increase in the number of doses as shown in Fig.\ \ref{coildoseno},
due to the decrease in the mass per dose, as can be seen from Fig.\ \ref{coilslope}. The volume of particles transported per dose 
in the system with 2 coils is more than that transported in the system with 8 coils. An increase in the number of
coils is not directly proportional to the output mass, i.e. a two-fold decrease in the number coils does not necessarily lead to
a two fold increase in the output dosed mass. For example, at the third dose, the configurations with 
2, 4 and 6 coils have cumulative dosed masses of 37grams, 24grams and  15grams, respectively. The under-prediction of the mass per dose is more extreme for less numbers of coils (or wider pitch)
but is close to the mass per dose for the simulation with the narrower pitch (7 to 8 coils).

\subsection{Locally averaged macroscopic velocity field}
\label{sec:locfields}

One advantage of performing simulation is the possibility for data-mining to obtain macroscopic fields from microscopic
data. In this section, we show the macroscopic velocity field that can be obtained from simulations.

\begin{figure} [!ht]
\centering
\includegraphics[scale=0.60, angle=270]{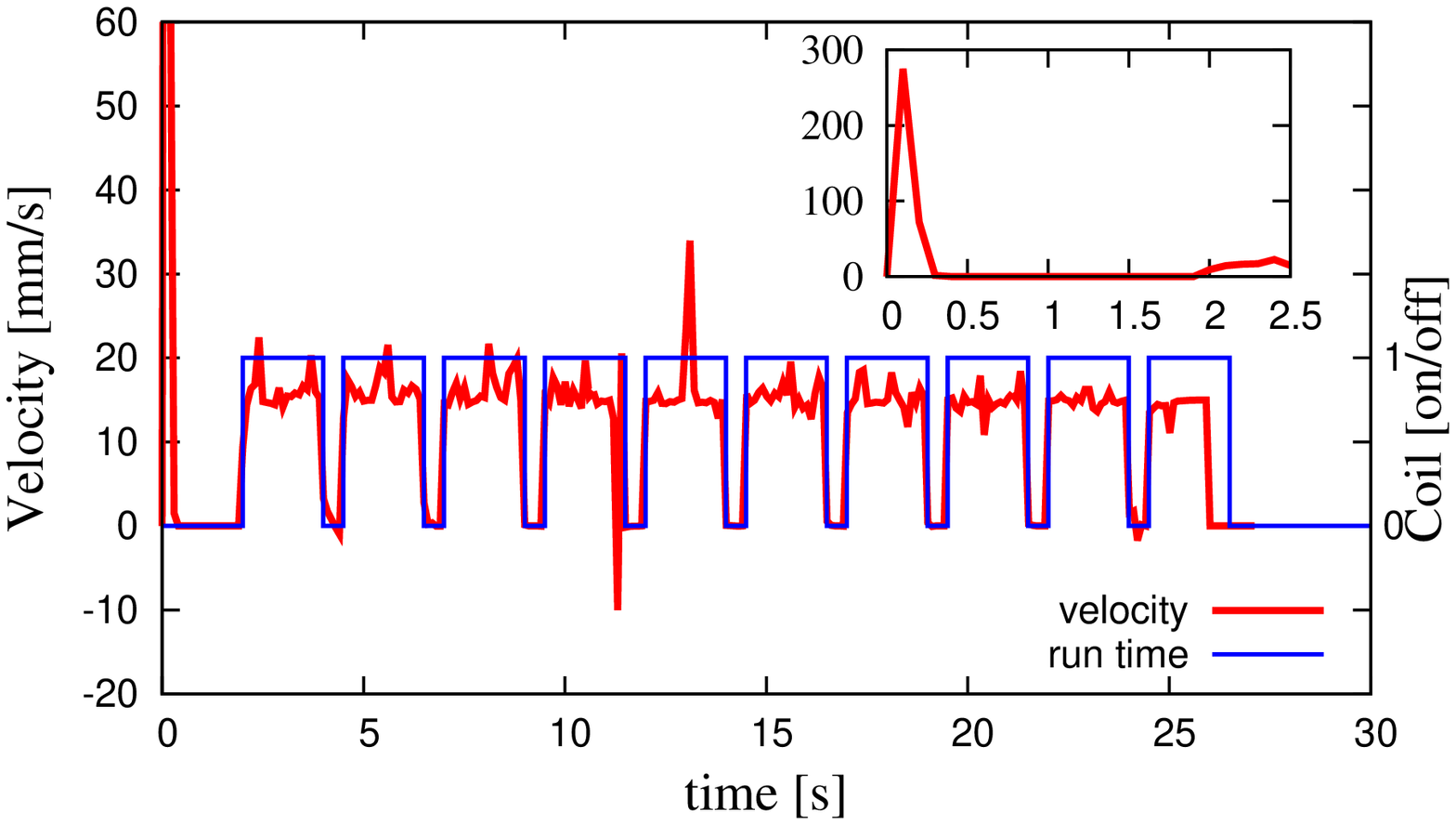}
\caption{Outlet material velocity (along $z$) during multiple dosing. The solid line represent the on-off motion of the coil. The coil angular velocity $\omega = $ 9.42 rad/s, $V_t = 94.5$ mm/s and $V_z = 2.09$ mm/s.}
\label{velstep}
\end{figure}

In Fig.\ \ref{velstep}, we show the velocity of the particles in the outlet of the box along with the staggered
motion of the coil (on = 1/off = 0). For this simulation,  the motion of the coil is such that an initial relaxation of 2 s allows the particles to 
settle during particle generation,  
followed by a staggered dosing phase where the coil is rotated  for 2 seconds at 90rpm, with waiting time of
0.5 seconds between successive doses until the box becomes empty. 

During the initial phase, where the particles
are, a momentary increase in the velocity can be seen as the particle fall to the base of the box and some escape through the outlet. 
% This is due to a few number of particles 
The particles quickly settle and the velocity drops to zero. Once the coil begins to move
at $t=$ 2 s, the velocity increases again with fluctuations and reaching a peak of $\approx$ 18 mm/s before 
steadily decreasing to zero as the coil motion goes to zero. The same pattern can be seen from the subsequent doses. It should
be noted that even though the velocity profile is mostly positive, we observe in some cases a negative velocity arising from a 
single particle moving in the opposite direction. This happens mostly during the 
dosing phase when the coil is moving and is possibly due to collision with other particles or due to violent contact with the boundary (coil).

\section{Conclusion}
\label{sec:doseconclusion}
%decreases non-linearly with
The dosing of cohesive powders in a simplified canister geometry has been studied using experiments and discrete 
element simulations. This work has highlighted the prospects of using discrete element simulations to model a
complex application test relevant in the food industry. While the modelling of cohesion remains a challenging issue, this
work highlights important aspects that can be useful for future research on this subject.

\begin{itemize}
 \item[i.] Scaling or coarse-graining of meso-particles by increasing their size relative to the primary (real) particles and 
setting appropriate parameters, e.g. timescales, to mimic the experimental particles makes it possible to simulate fine powders.

\item[ii.] Calibration of the interparticle friction and cohesive model parameters to match the experimental dosed mass leads to 
parameter values, different from those expected for the primary particles.

\item[iii.] Homogenization techniques to obtain macroscopic fields provides further insights into the dosing mechanisms beyond experimental methods.
\end{itemize}

Using the dosed mass as a target variable, we have shown experimentally
that the number of doses  shows an inverse proportionality to increasing dosage time. Consequently, the 
dosed mass shows a linear increase with dosage time as expected from the estimated mass per dose. Increasing 
the number of turns in the coil leads to a non-proportional increase of the dose mass. The mass output from the canister is found to 
show only a tiny sensitivity to the initial mass in the canister. 

All these observations have been confirmed by discrete element simulations for smaller masses while effects of arching and blockage were observed for higher masses are overestimated.
Future work will focus on the quantitative comparison for the masses as used in the experiments. An 
extraction of other macroscopic fields like density, stress or structure using homogenization tools can shed further light on the dosing process.
%The effects of wall friction and rolling resistance are presently being studied.

This work shows that scaling up particle diameter by more than ten times and adapting particle properties is a viable approach to overcome the 
untreatable number of particles inherent in experiments with fine, cohesive powders. The confidence gained in this study, focusing on a simplified canister geometry, 
paves the way to simulating the flow of these materials in more complex, real canister geometries, and to ultimately using numerical simulations as a virtual prototyping tool.

\section{Appendix A: Sensitivity studies on other parameters}
\label{sec:appendix}

\begin{figure} [!ht]
\centering
\includegraphics[scale=0.60, angle=270]{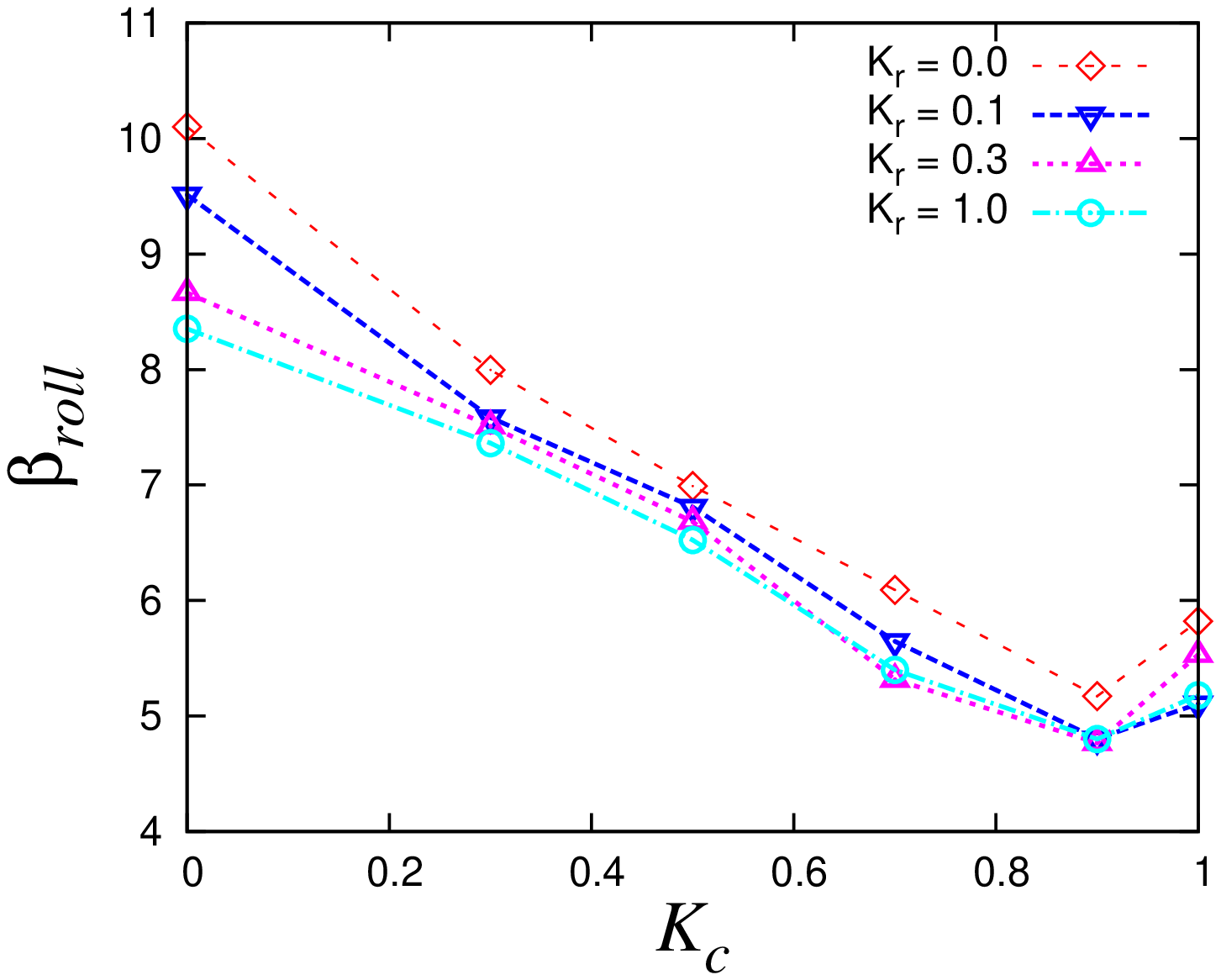}
\caption{Calibration of the cohesive stiffness $\bkc= \kc/k$ and rolling stiffness $\bkr = \kr/k$. We plot the mass per dose $\broll$ for different $\bkc$ and different $\bkr$ as
given in the inset.}
\label{roll_slope}
\end{figure}

In Fig.\ \ref{roll_slope}, we perform sensitivity studies on the effect of the rolling stiffness $\bkr = \kr/k$ with cohesive stiffness $\bkc= \kc/k$. We perform several simulation where the rolling stiffness
is fixed and cohesion is varied from 0 to 1. All other quantities are set according to Table\ \ref{parametertableb}. %Note that in all cases, the interparticle friction is set at $\mu$ 
For each simulation, the slope $\broll$ of the cumulative dosed mass obtained is plotted as function of the cohesive stiffness. 

Two observations are evident from Fig.\ \ref{roll_slope}. Firstly, we observe that $\broll$ decreases with increasing $\bkc$ leading to a slower/longer dosing process. At the highest $\bkc$, the 
cumulative dosed mass is no longer linear. At this point, the $\broll$ values obtained appear slightly higher since the dosage is uneven and an accurate measurement of $\broll$ is challenging. For much higher
cohesion (up to $\bkc =$ 10, not shown), no material flows out of the box.

%%%%%%%%%%%%%%%%%%%%%%%%

%%%%%%%%%%%%%%%%%%%%%%%%%%

\begin{figure} [!ht]
\centering
\includegraphics[scale=0.60, angle=270]{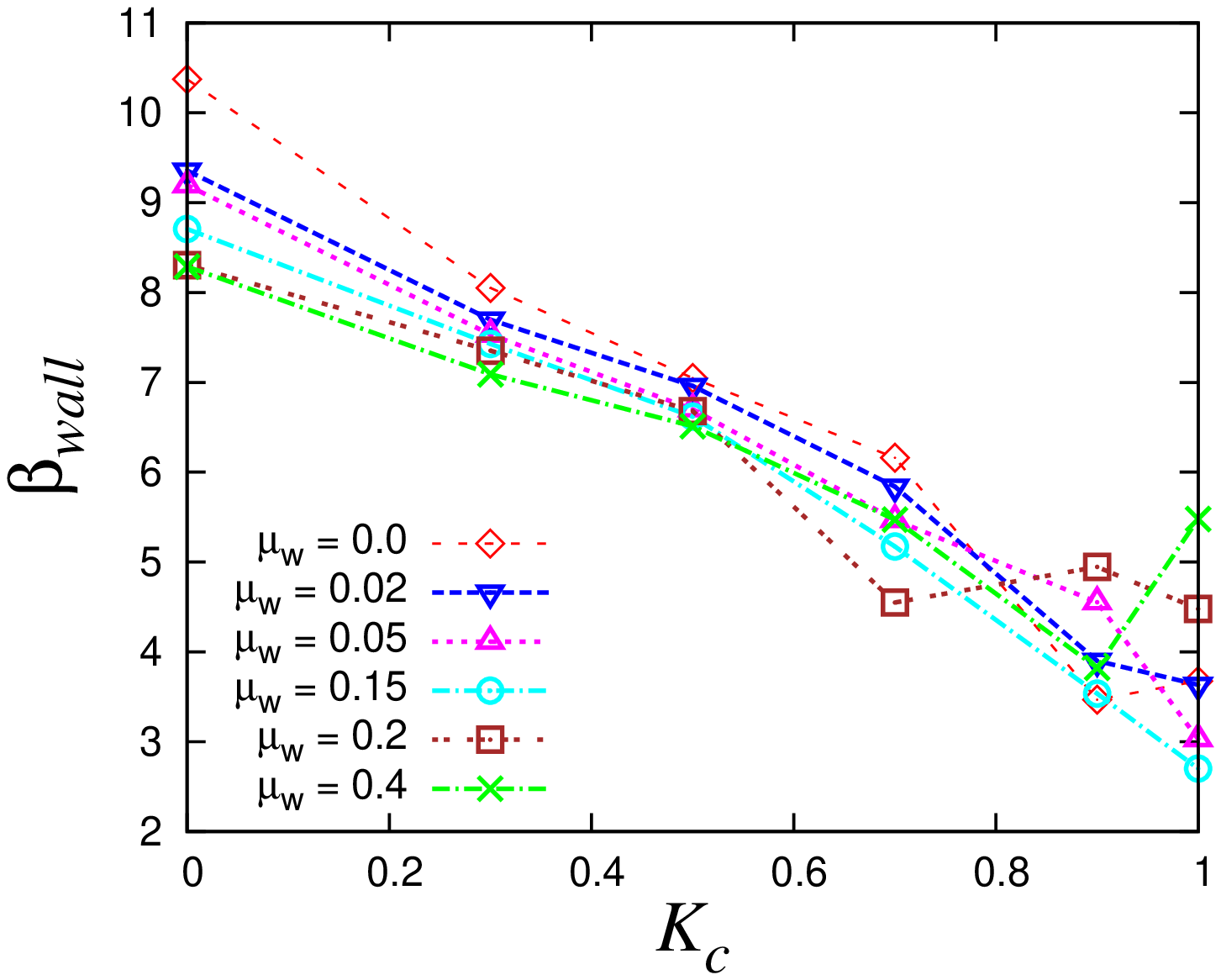}
\caption{Calibration of the cohesive stiffness $\bkc= \kc/k$ and wall friction $\mwall$. We plot the mass per dose $\broll$ for different $\bkc$ and different $\mwall$ as
given in the inset.}
\label{wall_slope}
\end{figure}
 
Secondly, $\broll$ is also found to decrease with increasing $\bkr$. The decrease is most visible in the cohesionless case where $\bkc = 0$ while $\broll$  is not much affected for cases with non-zero 
cohesion and rolling stiffnesses. In other words, $\bkr$ plays a minimal role in affecting the mass per dose $\broll$ -- in comparison to cohesive stiffness which, leads to a significant decrease
in the mass per dose.

In Fig.\ \ref{wall_slope}, we also consider how changes in the wall friction $\mwall$ and cohesive stiffness $\bkc= \kc/k$ affect the mass per dose $\bwall$. With increasing $\bkc$, the mass per dose 
$\bwall$ decreases, except for $\bkc = 1$ where uneven, nonlinear dosage leads to challenges in measuring $\bwall$. Increasing wall friction leads to a tiny decrease in $\bwall$. Due to this, we conclude that 
$\bkc$ is the most important quantity in determining the mass per dose.

%\end{linenumbers}

\section*{Acknowledgement}
Helpful discussions with A. Thornton, N. Kumar and M. Wojtkowski are gratefully acknowledged. 
This work is financially supported by the European Union funded 
Marie Curie Initial Training Network, FP7 (ITN-238577),
see {http://www.pardem.eu/} for more information.



% \newpage
% \bibliographystyle{elsarticle-harv} 
% \bibliography{references}

% \end{thebibliography}
%%%%%%%%%%%%%%%%%%%%%%%%%%%%%%%%%%%%%%%%

%\newpage
%%%%%%%%%%%%%%%%%%%%%%%%%%%%%%%%%

%%%%%%%%%%%%%%%%%%%%%%%%%%%%%%%%%%%%%%%%%%%%%%%%%%%%%%%%%%%%%%%%%%%%%%%%%%%%%%

%%%%%%%%%%%%%%%%%%%%%%%%%%%%%

%%%%%%%%%%%%%%%%%%%%%

%\newpage

%%%%%%%%%%%%%%%%%%%%%%%%%%%%%%%%%%%%%%%%%%

%%%%%%%%%%%%%%%%%%%%%%%%%%%%%%%%%%%%%%%%%%%%%%%%%%%%%%%%%%%

%%%%%%%%%%%%%%%%%%%%%%%%%%

%%%%%%%%%%%%%%%%%%%%%%%%%%%%%%%%%%%%%%%%%%%%%%%%%%%%%%%%%%%%%%%%%%%%%%%%%%%%%%%%%%%%%%%%%%%%%%

\end{document}